\documentclass{JFM-FLM_Au}

\makeatletter

\newcommand{\jfmplainpage}{\cppagefont\thepage}

\def\ps@headings{%
  \let\@mkboth\markboth
  \def\@oddhead{\hfill{\itshape\@righttitle}\hfill}%
  \def\@evenhead{\hfill{\itshape\@lefttitle}\hfill}%
  \def\@oddfoot{\hbox to \textwidth{\hfill\jfmplainpage}}%
  \def\@evenfoot{\hbox to \textwidth{\jfmplainpage\hfill}}%
  \def\sectionmark##1{\markboth{##1}{}}%
  \def\subsectionmark##1{\markright{##1}}%
}

\def\ps@myheadings{%
  \let\@mkboth\@gobbletwo
  \def\@oddhead{\hfil{\itshape\rightmark}\hfil\llap{\thepage}}%
  \def\@evenhead{\rlap{\thepage}\hfil\itshape\leftmark\hfil}%
  \def\@oddfoot{\hbox to \textwidth{\hfill\jfmplainpage}}%
  \def\@evenfoot{\hbox to \textwidth{\jfmplainpage\hfill}}%
  \def\sectionmark##1{}%
  \def\subsectionmark##1{}%
}

\def\ps@titlepage{%
  \leftskip\z@\let\@mkboth\@gobbletwo\vfuzz=5\p@
  \def\@oddhead{%
    \vbox{%
      \vspace*{-4pt}%
      \hbox to \textwidth{\@j@urnal \hfill}%
      \par\vskip4pt%
      \hbox to \textwidth{%
        {\fboxsep0pt\fbox{%
          \parbox{\textwidth}{%
            \par\vskip2.7pc
            \centerline{Banner appropriate to article type will appear here in typeset article}%
            \par\vskip2.7pc
          }%
        }}%
      }%
    }%
  }%
  \let\@evenhead\@oddhead
  \def\@oddfoot{\hbox to \textwidth{\hfill\jfmplainpage}}%
  \def\@evenfoot{\hbox to \textwidth{\jfmplainpage\hfill}}%
  \def\sectionmark##1{}%
  \def\subsectionmark##1{}%
}

\makeatother

\usepackage[version=4]{mhchem}
\usepackage{multirow}
\usepackage{cancel}
\usepackage{cleveref}

\lefttitle{ }
\righttitle{ }

\title{Acoustic radiation of thermodiffusively unstable turbulent lean premixed hydrogen--air flames}

\author{Francesco G. Schiavone\aff{1}, Guillaume Daviller\aff{2} \and Davide Laera\aff{1,3}}

\affiliation{\aff{1}Department of Mechanics, Mathematics and Management, Polytechnic University of Bari, Via Orabona 4, Bari 70125, Italy
\aff{2}CERFACS, 42 avenue Gaspard Coriolis, Toulouse 31057, France
\aff{3}Institut de Mécanique des Fluides de Toulouse, IMFT, Université de Toulouse, CNRS, Toulouse 31400, France}

\corresau{Davide Laera, \email{davide.laera@poliba.it}}

\begin{document}
\maketitle

\begin{abstract}
The impact of thermodiffusive effects on combustion noise in turbulent premixed slot jet flames is investigated using Direct Numerical Simulations.
Two thermodiffusively unstable lean hydrogen--air flames are compared with a thermodiffusively stable stoichiometric methane--air flame with comparable laminar properties and same turbulence intensity.
The hydrogen cases differ in bulk velocity, chosen to match either the turbulent flame brush length or the bulk velocity of the methane case.
Thermodiffusive effects are found to strongly alter both the heat release rate fluctuations, which dominate the far-field acoustic radiation, and the flame surface dynamics.
A theoretical framework extending the classical flamelet theory to thermodiffusively unstable flames is proposed and validated, relating the flame-generated sound to the time derivative of the flame surface area and to the stretch factor $I_0$.
The analysis identifies flame stretch as a key mechanism promoting noise radiation in thermodiffusively unstable flames.
Spectral analyses further show that hydrogen flames exhibit stronger low-frequency heat release rate fluctuations and reduced high-frequency content relative to the methane flame.
This is shown to be related to the coupled action of turbulence and thermodiffusive instabilities, which enhance large-scale flame motions while attenuating small-scale flame annihilation events.
Consequently, hydrogen flames radiate more strongly at low frequencies, near the acoustic peak, and exhibit a steeper high-frequency spectral roll-off.
Finally, Spectral Proper Orthogonal Decomposition reveals that hydrogen non-equidiffusion intensifies shear layer instabilities between combustion products and ambient air. These results indicate that thermodiffusive effects influence both direct and indirect combustion noise generation mechanisms in hydrogen flames.
\end{abstract}

\hrule{\hfill}

\section{Introduction}
\label{sec:intro}

Hydrogen (\ce{H2}) combustion holds significant potential to reduce the carbon emissions of gas turbines for aircraft propulsion and power generation~\citep{zhou2024hydrogen}. However, its peculiar properties ({\it e.g.} lower volumetric energy density, higher laminar burning velocity, higher adiabatic flame temperature, and higher molecular diffusivity) pose relevant scientific and technological challenges~\citep{pitsch2024transition}.
To lower the flame temperature and, consequently, the emissions of nitrogen oxides, lean premixed combustion is preferred~\citep{brewster1999modeling}.
For \ce{H2} flames, however, this condition promotes thermodiffusive (TD) effects and instabilities, given the resulting sub-unity Lewis number $Le$ of the reacting mixture~\citep{lapenna2023hydrogen}.
Moreover, lean premixed flames are more prone to thermoacoustic instabilities, arising from the resonant coupling between unsteady combustion and acoustic waves, which can possibly induce irreversible damage to combustion systems~\citep{poinsot2017prediction}.

Thermoacoustic instabilities can be triggered by the acoustic radiation of flames, {\it i.e.} combustion noise~\citep{dowling2015combustion,poinsot2017prediction}, which is emerging as an important issue also due to increasingly stringent noise regulations~\citep{kumar2025spectral}.
Combustion noise is conventionally distinguished into two components: direct and indirect.
Direct noise is related to the unsteady expansion and contraction induced by heat release rate (HRR) fluctuations~\citep{thomas1966flame,hurle1968sound,strahle1978combustion}.
It is broadband~\citep{tam2019combustion} and monopolar~\citep{ihme2017combustion}, and occurs even in unbounded spaces~\citep{dowling2015combustion}.
Indirect noise, on the other hand, stems from dipole sources related to the acceleration of non-uniform entropy, vorticity, or composition distributions~\citep{marble1977acoustic,magri2016compositional, hirschberg2022experimental,gentil2024, gentil2025}, as it occurs through the outlet nozzle of a combustor~\citep{dowling2015combustion}.

Heat release rate fluctuations are considered as the dominant noise source for unconfined turbulent premixed flames at low Mach number~\citep{strahle1985more,rajaram2009acoustic,swaminathan2011heat,haghiri2018sound,brouzet2021impact,ha2026direct}.
Under the assumption of an ideal gas mixture with constant average molecular weight and of an acoustically compact flame, the pressure fluctuation $p'$ in the far field at time $t$ and position $\boldsymbol{x}$ with respect to the flame is given by~\citep{dowling1992thermoacoustic,dowling2015combustion,brouzet2021impact,ha2026direct}:
\begin{equation}
    p'(\boldsymbol{x},t)=\frac{\gamma-1}{4\pi c_\infty^2\left|\boldsymbol{x}\right|}\left.\frac{\text{d}q}{\text{d}t}\right|_{t-\left|\boldsymbol{x}\right|/c_\infty},
    \label{eq:noise_hrr}
\end{equation}
where $\gamma$ is the heat capacity ratio, $c_\infty$ is the speed of sound in the far field medium, $q=\int_V\dot{\omega}_TdV$ is the integral of the local volumetric HRR $\dot{\omega}_T$, and $V$ is a compact volume.
This solution highlights the proportionality between the acoustic pressure perturbation and the time derivative of the HRR evaluated at a retarded time~\citep{dowling2015combustion}.
While valid in the acoustic far field, \Cref{eq:noise_hrr} does not hold true in the near field, which can be relevant for the onset of thermoacoustic instabilities inside a combustion chamber~\citep{pausch2019noise}.
In this sense, indirect noise sources, related to local variations of entropy gradients and flow velocity, as well as to turbulent structures in the burnt gases or interacting with the domain boundaries, may become dominant over HRR fluctuations, even in an unconfined space~\citep{talei2014comparative,pausch2019noise,ho2026sound}.

Based on flamelet theory~\citep{thomas1966flame,abugov1978acoustic,clavin1991turbulent},
a relation can be established between the rates of change of the global HRR and of the flame surface area, or, by considering local quantities, between the rates of change of the local volumetric HRR and of the flame surface density $\Sigma$~\citep{candel2009flame}.
A transport equation can be written for this latter quantity with the following general form~\citep{candel1990flame}:
\begin{equation}
    \frac{\text{d}\Sigma}{\text{d}t}=\epsilon\Sigma-\beta\Sigma^2,
    \label{eq:fsd_budget}
\end{equation}
indicating that $\Sigma$ increases when the flame is positively stretched ($\epsilon>0$), but with a limiting term ($\beta>0$).
Therefore, surface destruction mechanisms become dominant at larger deformations, and rapid flame annihilation phenomena should be the governing mechanism of generation of unsteady HRR fluctuations and of combustion noise~\citep{candel2009flame}.
This result was confirmed both experimentally~\citep{kidin1984sound,schuller2003self,candel2004flame} and numerically~\citep{talei2011sound,talei2014comparative,brouzet2019annihilation} for hydrocarbon flames, not affected by TD effects ($Le\approx 1$).
In particular, for these flames, a strong noise generation mechanism was identified in the detachment and subsequent consumption of unburnt gas pockets at the flame tip~\citep{schuller2003self,haghiri2018sound}, where most of the acoustic sources are concentrated~\citep{rajaram2009acoustic,brouzet2021impact}.

For lean premixed \ce{H2}--air flames, however, $Le<1$ and stretch impacts on the flame speed.
As established by the seminal study of~\citet{markstein1951experimental} and by later asymptotic analyses~\citep{Pelce_Clavin_1982,matalon1982flames,clavin1982effects,frankel1982effect}, these flames accelerate (decelerate) under positive (negative) stretch.
This stretch sensitivity induces intrinsic TD instabilities and cellular structures, thereby intensifying the chaotic flame front wrinkling and unsteady HRR fluctuations~\citep{lapenna2023hydrogen}.
In turbulent configurations, synergistic interactions are present between TD instabilities and turbulence, leading to a more accentuated wrinkling of the flame surface and to strong variations of the local reactivity~\citep{berger2022synergistic,lapenna2024synergistic}.
This favours the positive feedback between flame speed and stretch, which enhances the flame surface generation term in \Cref{eq:fsd_budget}, thus making flame surface destruction less dominant on the overall variation of the flame surface area and, consequently, of HRR~\citep{trouve1994evolution,han2008roles,chakraborty2011effects}.
Moreover, TD instabilities enhance the turbulent flame speed, leading to a reduction in the length of jet flames~\citep{berger2022synergistic}, to which the peak frequency of the acoustic spectrum is related~\citep{rajaram2009acoustic,liu2015modelling}.
As a consequence, for sub-unity $Le$ flames, an enhancement and a shift towards higher frequencies of flame-acoustics interactions is expected, as observed in experimental studies of thermoacoustic instabilities~\citep{yoon2017effects,lee2020combustion,paniez2024high}.

Previous one-dimensional (1-D)~\citep{talei2012parametric,jimenez2015sound} and two-dimensional (2-D)~\citep{talei2013direct} numerical analyses, however, indicated that, when $Le<1$, flames display reduced pressure fluctuations despite enhanced flame surface perturbations. This was linked to a slower and more progressive annihilation process driven by the gradual decline in the local consumption speed.
Although the reduced contribution of the destruction term is consistent with the stretch‑induced behaviour of sub‑unity $Le$ flames, it must be emphasised that these prior studies, confined to idealised 1-D and \mbox{2-D} laminar configurations, are insufficient to elucidate the influence of TD instabilities on combustion noise.
In this sense,~\citet{pedro2026impact} recently investigated the acoustics of 2-D thermodiffusively unstable laminar lean premixed \ce{H2}--air slit flames, and found that TD instabilities lead to sustained HRR oscillations that actively shaped combustion noise, with a trace in the acoustic spectra.
Nevertheless, to the best of our knowledge, sound generated by three‑dimensional (3-D) lean premixed turbulent \ce{H2}--air flames, for which TD instabilities can strongly affect the flame dynamics, has received limited attention in the literature~\citep{shoji2020effects,pillai2022investigation}. 
Therefore, the interplay between TD instabilities, turbulent mixing, unsteady HRR and acoustic radiation remains unclear.

Furthermore, it is known that coherent structures arise in unconfined non‑reacting turbulent jets~\citep{pickering2020lift} and dominate the acoustic radiation~\citep{jordan2013wave}.
\citet{brouzet2020role} reported that these structures likewise modulate the combustion noise of turbulent premixed methane--air flames in an unconfined ambient of equilibrium products, through their impact on the deformation of the flame front.
Moreover,~\citet{casel2022resolvent} established the relevance of wavepackets associated with the Kelvin--Helmholtz (K-H) shear layer instability at the interface between combustion products and a colder ambient in a swirl-stabilised turbulent premixed methane--air flame. Replacing methane (\ce{CH4}) with \ce{H2} yields combustion products of lower density and higher sound speed compared to pure air~\citep{nedden2025burner}, inevitably modifying the mixing and shear layer dynamics of an unconfined reacting jet in ambient air.
This is expected to impact the low frequency dynamics of lean premixed \ce{H2}--air flames~\citep{von2025low} and, consequently, the acoustic radiation.

The present work aims to address this subject by examining the acoustic radiation of 3-D lean  premixed \ce{H2}--air slot jet flames in an open environment of ambient air using Direct Numerical Simulations (DNS).
A focus is given to the role of TD instabilities on the noise generated due to the unsteady HRR, as well as to the relevance of the variation of the burnt gas properties on the generation of sound.
The characteristics of \ce{H2}--air flames are assessed by comparison with a thermodiffusively stable stoichiometric \ce{CH4}--air flame with similar unstretched laminar properties and identical turbulence intensity. 

The remainder of the article is organised as follows.
\Cref{sec:methodology} describes the slot jet flame configuration and numerical methodology considered in this study.
\Cref{sec:modelling} details the novel theoretical framework derived to relate TD effects to combustion noise.
Subsequently, the DNS results are discussed in~\Cref{sec:results} to characterise the flame dynamics and the corresponding acoustic radiation.
Then,~\Cref{sec:td_effects_noise} analyses how the impact of TD effects on the flame and jet dynamics affects the acoustic radiation, while~\Cref{sec:spod_noise} investigates the noise generation mechanisms using Spectral  Proper Orthogonal Decomposition (SPOD).
Finally,~\Cref{sec:conclusion} provides a summary of the main findings of this study.

\section{Numerical methodology}
\label{sec:methodology}
\subsection{Flame configurations and set-up}
\label{sec:setup}
\begin{figure}
    \centering
    \includegraphics[width=0.6\linewidth]{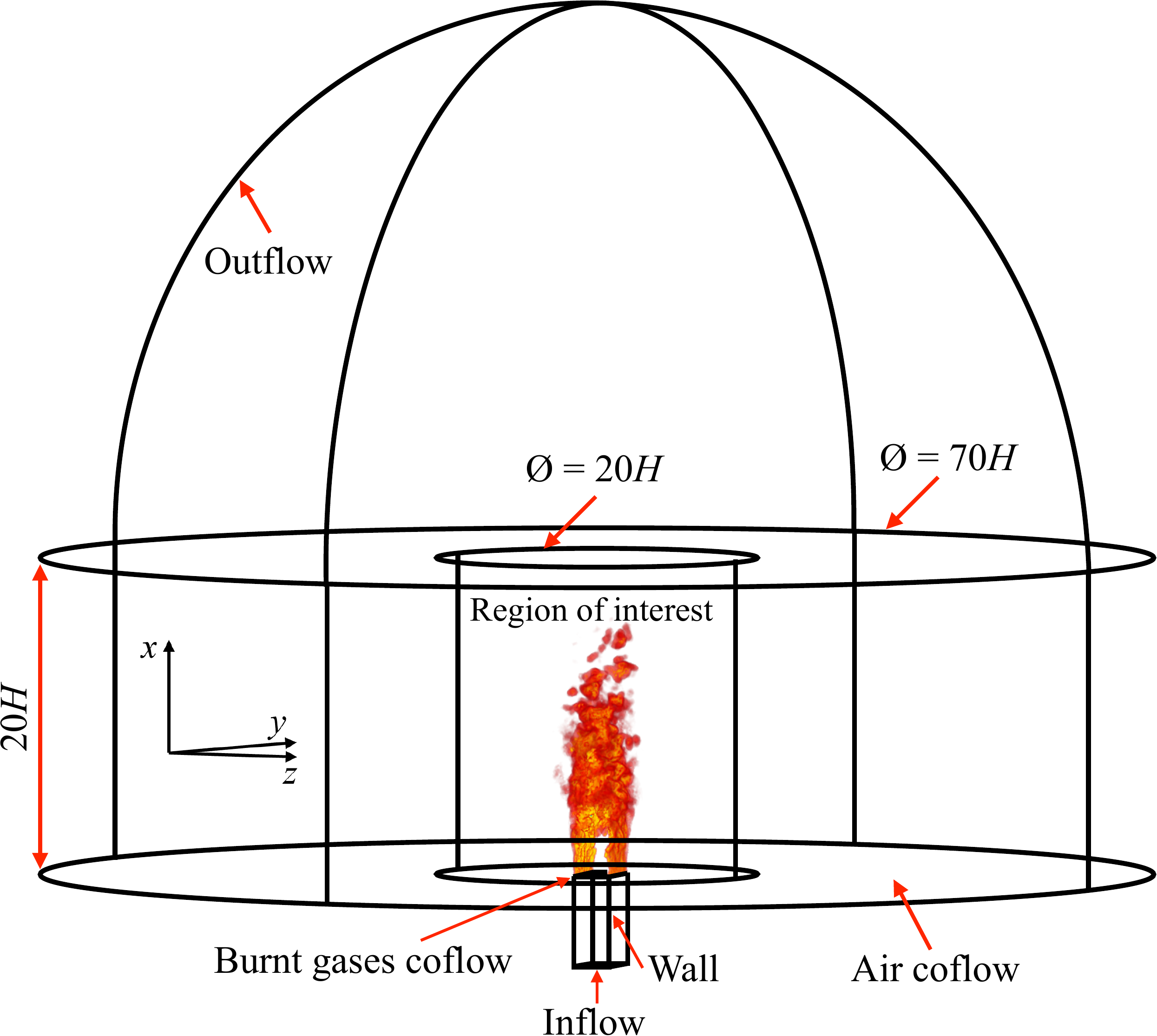}
    \caption{Schematic view of the computational domain (not to scale).}
    \label{fig:domain}
\end{figure}

The considered configuration features turbulent premixed slot jet flames in an open environment of ambient air at atmospheric pressure $p_{\infty}=1$ atm (the subscript ${\infty}$ denotes the far field medium conditions), similar to the one adopted by~\citet{schlimpert2016hydrodynamic,schlimpert2017analysis} to investigate combustion noise of \ce{CH4} flames.
A schematic representation of the 3-D computational domain is reported in~\Cref{fig:domain}.

Fresh turbulent gases with bulk velocity $U_B$ and temperature $T_u=300$~K (the subscript $u$ denotes the unburnt mixture conditions) are supplied through a central rectangular channel of height $H = 8.5$~mm in the cross-stream ($y$) direction, width $1.5H$ in the span-wise ($z$) direction, and stream-wise ($x$) length $3H$.
The jet is discharged into an open ambient air domain, represented by a cylinder of radius $35H$ with respect to the slot centre and span-wise length $L_x = 20H$, topped by a hemispherical cap.
Two laminar coflows of equilibrium burnt gases at the adiabatic flame temperature are injected in the span-wise direction along the major sides of the channel outlet to sustain the flame, following analogous configurations in the literature~\citep{schlimpert2016hydrodynamic,schlimpert2017analysis,lapeyre2019training,coulon2023direct,male2025hydrogen}.
Each has a cross‑stream width equal to $H/12$ and an axial velocity $U_{b,cf} = 0.01U_B$ (the subscript $b$ denotes the burnt gases conditions).
All is surrounded by a laminar ambient air coflow with axial velocity $U_{air} = 0.1$~m~s$^{-1}$, with smooth inter-flow transitions imposed via a hyperbolic tangent function.
This marks a difference with respect to most of the studies in the literature performed in similar jet flame configurations~\citep{haghiri2018sound,lapeyre2019training,brouzet2021impact,coulon2023direct,male2025hydrogen}, which consider a burnt gases environment.

\begin{table}
  \begin{center}
    \def~{\hphantom{0}}
    \begin{tabular}{lccccc}
      Fuel & $\phi$ & $S_L^0$ [m s$^{-1}$] & $\delta_L^0$ [mm] & $T_{ad}$ [K] & $\dot{\omega}_{T,max}^{1D}$ [W m$^{-3}$] \\[3pt]
      \ce{CH4} & $1.0$ & 0.38 & 0.40 & 2260 & $4.60\times 10^9$\\  
      \ce{H2}  & $0.45$ & 0.38 & 0.45 & 1535 & $1.50\times 10^9$\\
    \end{tabular}
    \caption{Laminar flame properties of the two reacting mixtures considered in this work.}
    \label{tab:lam_properties}
  \end{center}
\end{table}

Two fuel--air mixtures are considered in this work. Their laminar flame properties, derived from 1-D unstretched laminar premixed flame computations in Cantera~\citep{goodwin2023cantera}, are reported in~\Cref{tab:lam_properties}: the flame speed $S_L^0$, the thermal flame thickness $\delta_L^0=(T_b-T_u)/\left|\nabla T\right|$~\citep{jarosinski1984thickness}, the adiabatic flame temperature $T_{ad}$, and the maximum HRR $\displaystyle\dot{\omega}_{T,max}^{1D}$. 
Three turbulent flames are studied.
First, a stoichiometric \ce{CH4}--air flame with global equivalence ratio $\phi_g=1.0$ and bulk velocity $U_B=10$~m$~$s$^{-1}$, hereafter denominated M10, is considered as a reference case, since it is designed not to feature TD instabilities.
Subsequently, two lean ($\phi_g=0.45$) \ce{H2}--air flames are investigated: one (H25) with bulk velocity $U_B=25$~m$~$s$^{-1}$, and the other (H10) with bulk velocity $U_B=10$~m$~$s$^{-1}$, {\it i.e.} the same of the \ce{CH4} flame.
The former case is chosen to have a comparable average length of the turbulent flame brush with respect to the M10 case, since this dimension impacts the peak frequency of the acoustic spectrum~\citep{rajaram2009acoustic,liu2015modelling}, while the latter one is considered to further investigate the impact of the jet Reynolds number.

\Cref{tab:flow_params} provides relevant flow and flame parameters for the three cases. In particular, the thermal power is $P_{th}=Y_{F,u}\rho_uA_{ch}U_Bh_{lv}$, where $Y_{F}$ and $\rho$ are, respectively, the fuel mass fraction and the density, $A_{ch}=1.5H^2$ is the channel section, and $h_{lv}$ is the fuel lower heating value, equal to $5.0\times10^7$~J~kg$^{-1}$ for \ce{CH4} and to $1.2\times10^8$~J~kg$^{-1}$ for \ce{H2}.
The square jet Reynolds number is defined as $Re=(U_BH)/\nu_u$, where $\nu$ is the kinematic viscosity.
The inlet Mach number is $M=U_B/c_u$, with $c$ being the speed of sound, while $\tau=L_x/U_B$ is the flow through time.
The progress variable $C$ is defined as~\citep{male2025hydrogen}:
\begin{equation}
    C=\frac{Y_F-Y_{F,u}(\xi)}{Y_{F,b}(\xi)-Y_{F,u}(\xi)}=\frac{Y_F-\xi}{\max[0, (\xi-\xi_s)/(1-\xi_s)]-\xi},
    \label{eq:progvar}
\end{equation}
where $\xi$ is the mixture fraction defined by~\citet{bilger1989structure}, and $\xi_s$ is its stoichiometric value.
The average flame length $L_f$ is evaluated by considering the stream-wise extension of the mean flame sheet, defined by the Favre-averaged progress variable at $\widetilde{C}=C^*$, {\it i.e.} the value of $C$ at the location of the maximum HRR value in the corresponding 1-D laminar unstretched premixed flame (see~\Cref{tab:lam_properties}).
To isolate the reacting zone from the mixing region with cold ambient air, values of $\xi$ lower than $0.5 Y_{F,u}$ are not considered.
In analogy with the works by~\citet{luca2019statistics} and~\citet{berger2022synergistic}, the Kolmogorov length scale $\eta$ is estimated \textit{a posteriori} at $x=L_f/2$ on the previously defined mean flame sheet as $\displaystyle\eta=(\bar{\nu}^3/\tilde{\varepsilon})^{1/4}$, using the ensemble-averaged kinematic viscosity $\bar{\nu}$ and the Favre-averaged turbulence dissipation rate $\tilde{\varepsilon}$~\citep{pantano2003mixing}, with the averages performed in time and in the span-wise direction.
To allow for a better comparison between the different cases, featuring different thermal powers and turbulent flame characteristic dimensions, a reference acoustic pressure $p_{ref}$ is defined (see~Appendix~\ref{app:pressure}), based on \Cref{eq:noise_hrr} and on combustion noise scaling laws~\citep{rajaram2006premixed,candel2009flame}:
\begin{equation}
    p_{ref} = \frac{\gamma-1}{4\pi c_\infty^2}\frac{Y_{F,u}\rho_uHU_B^2h_{lv}}{L_f}.
    \label{eq:p_ref}
\end{equation}
Finally, the Karlovitz number is estimated as $Ka=(l_t/\delta_L^0)^{-1/2}(u'/S_L^0)^{3/2}$, where $l_t$ is the integral length scale of turbulence and $u'$ is the turbulent fluctuation intensity.
For all cases considered in this study, $u' = 2.5$~m~s$^{-1}$ and $l_t=H/4=2.1$~mm, in analogy with the previous work by~\citet{coulon2023direct}.

\begin{table}
  \begin{center}
    \def~{\hphantom{0}}
    \begin{tabular}{lcccccccccccc}
      Case & Fuel & $\phi_g$ & $U_B$ [m~s$^{-1}$] & $P_{th}$ [kW] & $Re$ & $M$ & $\tau$ [ms] & $L_f/H$ & $C^*$ & $\eta$ [$\mu$m] & $p_{ref}$ [Pa] & $Ka$\\[3pt]
      M10 & \ce{CH4} & 1.0 & 10 & 3.4 & 5~300 & 0.03 & 17.06 & 4.1 & 0.77 & 90 & 20.2 & 7 \\
      H25 & \ce{H2} & 0.45 & 25 & 4.3 & 10~830 & 0.07 & 6.824 & 4.3 & 0.80 & 40 & 59.6 & 8 \\
      H10 & \ce{H2} & 0.45 & 10 & 1.7 & 4~330 & 0.03 & 17.06 & 2.4 & 0.80 & 70 & 17.5 & 8 \\
    \end{tabular}
    \caption{Flow and flame parameters for the three DNS configurations.}
    \label{tab:flow_params}
  \end{center}
\end{table} 

\begin{figure}
  \centerline{\includegraphics[width=100mm]{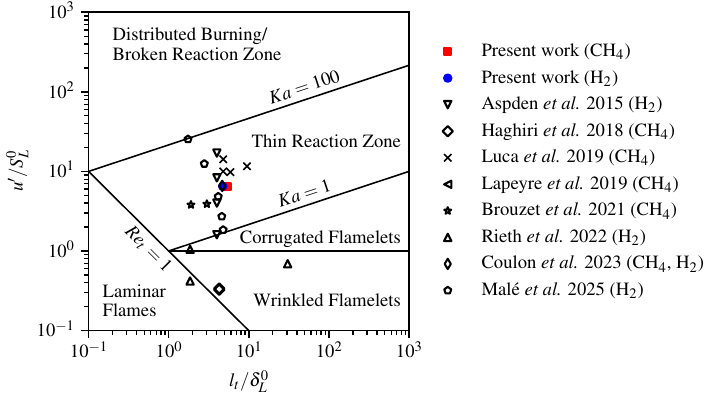}}
  \caption{Location in the Borghi--Peters turbulent combustion diagram~\citep{peters1999turbulent} of \ce{CH4} and \ce{H2} flames of the present study and of DNS in the literature~\citep{aspden2015turbulence,haghiri2018sound,luca2019statistics,lapeyre2019training,brouzet2021impact,rieth2022enhanced,coulon2023direct,male2025hydrogen}.}
  \label{fig:borghi}
\end{figure}

Based on the selected parameters, the flames investigated in this work lie in the Thin Reaction Zone of the Borghi--Peters turbulent combustion regime diagram~\citep{peters1999turbulent}.
This is presented in~\Cref{fig:borghi}, together with DNS data of premixed \ce{CH4} and \ce{H2} flames reported in the literature for comparison.
It should be noted that the regime diagram is here provided only for an order-of-magnitude analysis, and care should be taken in its interpretation~\citep{veynante2002turbulent,poinsot2005theoretical,skiba2018premixed}.
Nevertheless, for the premixed flames here considered, viscosity and density remain similar, due to the dominant influence of nitrogen dilution~\citep{coulon2023direct}. This, combined with the comparable turbulence fields (see Appendix~\ref{app:turbulence}) and laminar flame properties (see~\Cref{tab:lam_properties}), sustains the fact that the three cases fall within the same regime.

\subsection{Numerical approach}
\label{sec:num_chem}

The DNS in this work were performed using the explicit massively parallel compressible multi-species Navier--Stokes solver AVBP~\citep{schonfeld1999steady}, extensively adopted in the literature for DNS of turbulent reacting flows~\citep{lapeyre2019training,coulon2023direct,gaucherand2024dns,male2025hydrogen}.
The oxidation of \ce{H2} is described with the San Diego mechanism, featuring 9 species and 21 reactions~\citep{saxena_testing_2006}, while a semi-global two-steps mechanism~\citep{selle2004compressible}, involving 6 species, is adopted for \ce{CH4}.
The dynamic viscosity $\mu$ is approximated with a power law function.
Thermal diffusivity is computed from the viscosity by assuming a constant Prandtl number $Pr$ for the mixture, while species diffusivities are evaluated by considering a constant Schmidt number $Sc$, specific for each species (see Appendix~\ref{app:transport}).

The two-steps finite-element Taylor--Galerkin scheme TTGC, providing a third-order accuracy in space and time, is adopted to discretise convective terms, while a second-order finite-element Galerkin scheme is used for diffusion terms~\citep{colin_development_2000}.
No-slip adiabatic boundary conditions are applied at walls in the central rectangular channel (see~\Cref{fig:domain}). Inflows and outflows are treated with the Navier--Stokes Characteristic Boundary Conditions (NSCBC)~\citep{poinsot1992boundary}. 
To avoid non-physical acoustic reflection, the non-reflecting NSCBC formulation~\citep{daviller2019generalized} is adopted at the inlet, while a relaxation coefficient $K=100$ s$^{-1}$ is imposed at the outlet, leading to a cut-off frequency $f_c=K/(4\pi)\approx 8$~Hz~\citep{selle2004actual}.

The channel inlet is forced using homogeneous and isotropic turbulence, as in previous DNS in the literature~\citep{coulon2023direct,male2025hydrogen}. The turbulent velocity fluctuation $\boldsymbol{u'}$ is obtained with a synthetic generation method based on a Fourier series decomposition~\citep{kraichnan1970diffusion}:
\begin{equation}
    \boldsymbol{u'}(\boldsymbol{x}, t) = \sum_{n=1}^N\left[\hat{\boldsymbol{v}}^n\boldsymbol{k}^n\cos\left(\boldsymbol{k}^n\boldsymbol{x}+\omega^nt\right)+\hat{\boldsymbol{w}}^n\boldsymbol{k}^n\sin\left(\boldsymbol{k}^n\boldsymbol{x}+\omega^nt\right)\right],
\end{equation}
with $N=200$ modes.
The Fourier modes $\{\boldsymbol{v}^n,\boldsymbol{w}^n\}$, wave vectors $\boldsymbol{k}^n$ and pulsations $\omega^n$ are random variables sampled to obtain a Passot--Pouquet spectrum~\citep{passot1987numerical}:
\begin{equation}
    E(k)=16\frac{u'^2}{k_e}\sqrt{\frac{2}{\pi}}\left(\frac{k}{k_e}\right)^4\exp\left[-2\left(\frac{k}{k_e}\right)^2\right],
    \label{eq:passot_pouquet}
\end{equation}
where $k_e$ is the wave number associated with the most energetic eddies, related to the integral length scale of turbulence through $l_t$~$=$~$\sqrt{2\pi}/k_e$. A spectral analysis of the resulting turbulence field is provided in Appendix~\ref{app:turbulence}.

The computational domain in~\Cref{fig:domain} is discretised with an unstructured grid of tetrahedral elements.
The region of interest for the investigation of noise is represented by a cylinder of radius $10H$ with respect to the slot centre and length $L_x=20H$ from the channel exit (see~\Cref{fig:domain}).
Inside this region, the grid is sensibly refined to properly resolve the flame front and the turbulent structures, and to capture the acoustic waves.
The grid resolution in the flame region is not greater than $\Delta_x=100$~$\mu$m, in line with previous DNS studies performed for \ce{CH4}~\citep{lapeyre2019training} and \ce{H2}~\citep{male2025hydrogen} turbulent premixed flames under analogous operating conditions.
This resolution ensures at least 7-10 points in the flame front, and a reasonable time step $\Delta_t$, based on the acoustic Courant-Friedrichs-Lewy ($CFL$) number:
\begin{equation}
    \Delta_t = \frac{CFL\Delta_x}{\max(\left|u+c\right|,\left|u-c\right|)},
\end{equation}
with $CFL \leq 0.7$. This results in a grid size of approximately 210 million cells for the M10 and H25 cases, and 165 million cells for the H10 case, while $\Delta_t$ is equal to 19.7~ns, 23.3~ns and 24.7~ns for the M10, H25 and H10 cases, respectively.
A grid independence study was performed for the H25 case to verify that turbulent combustion statistics are not affected by further mesh refinements (see Appendix~\ref{app:mesh}).
To reduce the computational cost, a smooth grid stretching was applied outside the reacting region, with a growth factor of the cell sizes not greater than 2\% to avoid any spurious waves~\citep{haghiri2018sound,brouzet2021impact}.
Outside the region of interest, the grid was more strongly stretched to further avoid acoustic reflections from the outflow boundaries and dissipate high frequency spurious components of pressure.

The simulations were initialised with DNS flame solutions from~\citet{lapeyre2019training} for the \ce{CH4} flame, and from~\citet{coulon2023direct} for the \ce{H2} ones.
Statistical convergence was verified by analysing the cumulative temporal mean and r.m.s. statistics of the volume integral of HRR, of the streamwise velocity at the outlet of the rectangular channel, and of the pressure in the region far from the jet. After reaching convergence, production runs were performed for 4$\tau$, long enough to capture the peak of the combustion noise acoustic spectrum (see~\Cref{sec:acoustic}).
The M10 and H25 cases are representative of the minimum and maximum computational costs sustained for the considered cases.
For the M10 case, the cost of the production runs was approximately equal to 1.4 million CPU-hours, obtained using 40 nodes, each equipped with two AMD EPYC Rome 7H12 64-core processors.
The H25 case, instead, required approximately 700 thousand CPU-hours, running on 30 nodes.
In terms of physical wall-clock time, these corresponded to approximately 270 and 180 hours for the M10 and H25 cases, respectively.

\section{Modelling of the acoustic radiation}
\label{sec:modelling}

The acoustic radiation of turbulent premixed flames is commonly interpreted in terms of the unsteady evolution of the global HRR, as shown in \Cref{eq:noise_hrr}.
Coherently with the classical combustion noise flamelet theory~\citep{abugov1978acoustic,clavin1991turbulent}, fluctuations of the global HRR are governed primarily by the generation and destruction of the (turbulent) flame surface area $A_T$.
Therefore, for an equidiffusive mixture ($Le\approx 1$), for which the effect of stretch on the local velocity is not relevant and the Damk\"ohler's first hypothesis holds~\citep{damkohler1940einfluss,chakraborty2019validity}, the far‑field pressure fluctuation may be written as~\citep{candel2009flame}:
\begin{equation}
    p'(\boldsymbol{x},t)=\frac{\rho_\infty}{4\pi\left|\boldsymbol{x}\right|}\left(\frac{\rho_u}{\rho_b}-1\right)S_L^0\left.\frac{\text{d}A_T}{\text{d}t}\right|_{t-\left|\boldsymbol{x}\right|/c_\infty},
    \label{eq:p_area_le1}
\end{equation}
where $A_T$ is here evaluated numerically as~\citep{vervisch1995surface}:
\begin{equation}
    A_T=\int_0^1\Sigma(C^*)dC^*=\int_V\left|\nabla C\right|dV.
    \label{eq:turb_flame_surf}
\end{equation}
The expression in \Cref{eq:p_area_le1} has been used in several experimental studies examining the sound radiated by acoustically forced laminar flames \citep{schuller2002dynamics,schuller2003self} and by turbulent flames \citep{belliard1997etude,truffaut1998etude}.
It also formed the basis of the numerical analysis by \citet{brouzet2021impact} of the impact of stretch on turbulent premixed \ce{CH4}--air flame acoustics.

In the case of a non-equidiffusive mixture ($Le\neq 1$), however, the flame surface dynamics and reactivity are impacted by stretch effects.
In a general form, the flame--turbulence interaction can be described in this case by:
\begin{equation}
\frac{S_T}{S_L^0} = I_0 \frac{A_T}{A_0},
    \label{eq:speed_area}
\end{equation}
where $I_0$ is the stretch factor, $A_0$ is a reference flame surface, and $S_T$ is the (turbulent) flame consumption speed, defined as~\citep{poinsot2005theoretical}:
\begin{equation}
    S_T=-\frac{1}{\rho_u Y_{F,u}A_0}\int_V\dot{\omega}_FdV,
    \label{eq:turb_speed}
\end{equation}
where $\dot{\omega}_F$ is the local fuel reaction rate.
For mixtures featuring $Le < 1$, the local flame velocity is strongly affected by stretch, with $I_0 > 1$~\citep{berger2022synergistic,coulon2023direct}.
Therefore, the assumption underlying \Cref{eq:p_area_le1} no longer holds~\citep{trouve1994evolution,lapenna2023hydrogen}.
In this sense,~\citet{talei2013direct} applied \Cref{eq:p_area_le1} to acoustically forced 2-D laminar flames, with $S_L^0$ replaced by the time-averaged consumption speed to account for stretch effects, and found a systematic over-prediction of the sound emission.
This highlights the need for a more general formulation, to account explicitly for the influence of stretch on the local flame propagation speed.
A suitable starting point is the following expression~\citep{candel2009flame}:
\begin{equation}
    p'(\boldsymbol{x},t)=\frac{\rho_\infty}{4\pi\left|\boldsymbol{x}\right|}\left(\frac{\rho_u}{\rho_{b}}-1\right)\left.\frac{\text{d}}{\text{d}t}\int_{A_T}S_d\text{d} A\right|_{t-\left|\boldsymbol{x}\right|/c_\infty},
    \label{eq:p_area}
\end{equation}
where $S_d$ is the local flame displacement speed, dependent on both stretch and time.
\Cref{eq:p_area} is valid for arbitrary Lewis numbers and incorporates the effects of stretch on the flame dynamics through $S_d$. Within a finite thickness flame, the value of $S_d$ is highly sensitive to the definition of the flame surface~\citep{dave2020evolution}, {\it i.e.} to the selected isosurface $C=C^*$. To remove this ambiguity, in the present work $S_d$ is taken as the density‑weighted displacement speed $\displaystyle \left.(\rho S_d)\right|_{C=C^*}/\rho_u$ (see Appendix~\ref{app:stretch}).

Recalling \Cref{eq:speed_area} and using the relation between $S_d$ and $S_T$~\citep{dave2020evolution}, one obtains:
\begin{equation}
\int_{A_T}S_d\text{d} A=\underbrace{\frac{1}{A_0}\int_{A_T}S_d\text{d} A}_{S_T}A_0=S_TA_0=S_L^0I_0A_T,
\label{eq:SdSc}
\end{equation}
which shows the influence of stretch on the acoustic source term through $I_0$. Substituting \Cref{eq:SdSc} into \Cref{eq:p_area} yields:
\begin{equation}
p'(\boldsymbol{x},t)=\frac{\rho_\infty}{4\pi\left|\boldsymbol{x}\right|}\left(\frac{\rho_u}{\rho_{b}}-1\right)S_L^0\left.\frac{\text{d}(I_0A_T)}{\text{d}t}\right|_{t-\left|\boldsymbol{x}\right|/c_\infty}.
\label{eq:p_area_i0}
\end{equation}

\Cref{eq:p_area_i0} generalises the classical flamelet-based expression of \Cref{eq:p_area_le1} by introducing the stretch factor $I_0$.
This provides a unified description of the acoustic radiation of both equidiffusive and non‑equidiffusive turbulent premixed flames.
In particular, \Cref{eq:p_area_i0} shows that the combustion noise source term is governed by the temporal variation of the turbulent flame surface area, modulated by the influence of stretch.

Moreover, for a fixed operating condition, the global value of $I_0$ is often observed to vary weakly in time for a statistically steady state~\citep{luca2019statistics,attili2021turbulent,berger2022synergistic,coulon2023direct}. Under this assumption, \Cref{eq:p_area_i0} reduces to:
\begin{equation}
    p'(\boldsymbol{x},t)=\frac{\rho_\infty}{4\pi\left|\boldsymbol{x}\right|}\left(\frac{\rho_u}{\rho_{b}}-1\right)S_L^0\overline{I_0}\left.\frac{\text{d}A_T}{\text{d}t}\right|_{t-\left|\boldsymbol{x}\right|/c_\infty},
    \label{eq:p_area_i0_avg}
\end{equation}
where $\overline{I_0}$ is the time-averaged value of $I_0$.
\Cref{eq:p_area_i0_avg} highlights that non-equidiffusion effects amplify the acoustic emission by a factor $\overline{I_0}$, all else being equal.

\section{Analysis of DNS results}
\label{sec:results}

First, a detailed analysis of the flame dynamics, the flame--turbulence interaction and the acoustic field radiated by the three flames is provided hereafter.

\subsection{Flame--turbulence interaction}
\label{sec:flame_turbulence_int}

\begin{figure}
  \centerline{\includegraphics[width=90mm]{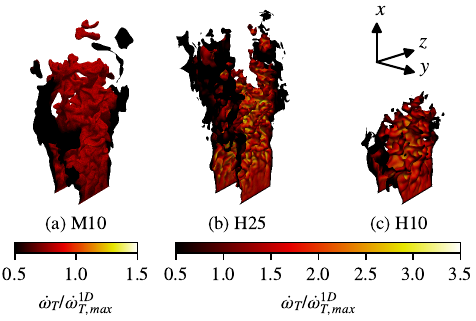}}
  \caption{Isosurfaces of progress variable $C=C^*$ coloured by the normalised HRR $\dot{\omega}_T/\dot{\omega}_{T,max}^{1D}$ taken at $t=4\tau$ for the three cases.}
  \label{fig:slot_isosurf}
\end{figure}

\Cref{fig:slot_isosurf} shows isosurfaces of progress variable taken at $C=C^*$, coloured by the normalised local HRR $\dot{\omega}_T/\dot{\omega}_{T,max}^{1D}$ (see~\Cref{tab:lam_properties}), for the three cases.
As expected, the \ce{CH4} flame has a regular, thermodiffusively stable behaviour, with the value of $\dot{\omega}_T$ constant on the whole flame surface.
The only exception is the outer region, where the mixing with ambient air decreases both the local temperature and equivalence ratio, hence affecting also $\dot{\omega}_{T}$. The \ce{H2} flames, in contrast, both display a highly corrugated front, together with pronounced local fluctuations of $\dot{\omega}_T$.
Furthermore, the H25 configuration yields a turbulent flame brush length comparable to that of the \ce{CH4} flame, consistent with the design objective for this case (see~\Cref{sec:intro}).
Conversely, the H10 case is characterised by a substantially shorter brush, despite sharing the same value of $U_B$ as the M10 one, thus reflecting the higher burning rate of thermodiffusively unstable \ce{H2} flames~\citep{berger2022synergistic}.
Overall, these results are consistent with previous studies in the literature on analogous configurations~\citep{coulon2023direct}.

\begin{figure}
  \centerline{\includegraphics[width=130mm]{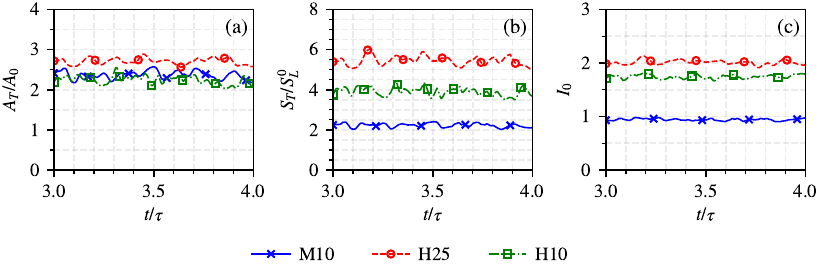}}
  \caption{Temporal evolution of the normalised turbulent flame surface $A_T/A_0$, of the normalised turbulent flame consumption speed $S_T/S_L^0$, and of the stretch factor $I_0$ for the three cases.}
  \label{fig:st_at_i0}
\end{figure}

The impact of flame--turbulence interaction on the global flame behaviour is highlighted in~\Cref{fig:st_at_i0}, which reports, for the three cases, the temporal evolution over one flow through time $\tau$ of the normalised turbulent flame surface $A_T/A_0$ (see \Cref{eq:turb_flame_surf}), of the normalised turbulent flame consumption speed $S_T/S_L^0$ (see \Cref{eq:turb_speed}), and of the stretch factor $I_0$ (see \Cref{eq:speed_area}). Here, the reference flame surface $A_0$ is taken as the area of the mean progress variable field at $\widetilde{C}=C^*$.
Considering $A_T/A_0$ first, the three flames exhibit broadly similar trends and magnitudes, consistent with their exposure to comparable turbulence fields (see Appendix~\ref{app:turbulence}).
This agreement is most evident for the M10 and H10 cases, which share the same bulk velocity, whereas the H25 flame shows a modest increase in $A_T/A_0$.
For lean premixed \ce{CH4}--air jet flames, \citet{attili2021turbulent} reported that, at approximately constant $Ka$, increasing $Re$ enhances turbulence within the inner flame layer, thereby increasing both $A_T/A_0$ and $S_T/S_L^0$.
In the present lean \ce{H2}--air flame, this effect is further amplified by TD instabilities, which interact synergistically with turbulence to promote flame surface generation.
This synergy also induces local equivalence ratio fluctuations, leading to variations in reactivity, HRR, and ultimately $S_T$~\citep{berger2022synergistic}. 

In this sense,~\Cref{fig:st_at_i0}(b) shows that, while both \ce{H2} flames display higher $S_T$ values than the thermodiffusively stable M10 case, the increase is stronger for the H25 case. 
As a result, the time-averaged value of the stretch factor $\overline{I_0}$, reported in~\Cref{fig:st_at_i0}(c), passes from 1.7 for the H10 case to 2.0 for the H25 one.
The value of $\overline{I_0}$ for the H25 case agrees with the DNS results of \citet{coulon2023direct} performed under analogous operating conditions, further supporting the present numerical configuration.
As expected, the M10 case yields $\overline{I_0}\approx 1$  (see~\Cref{fig:st_at_i0}(c)), consistently with the absence of TD effects.

\begin{figure}
  \centerline{\includegraphics[width=130mm]{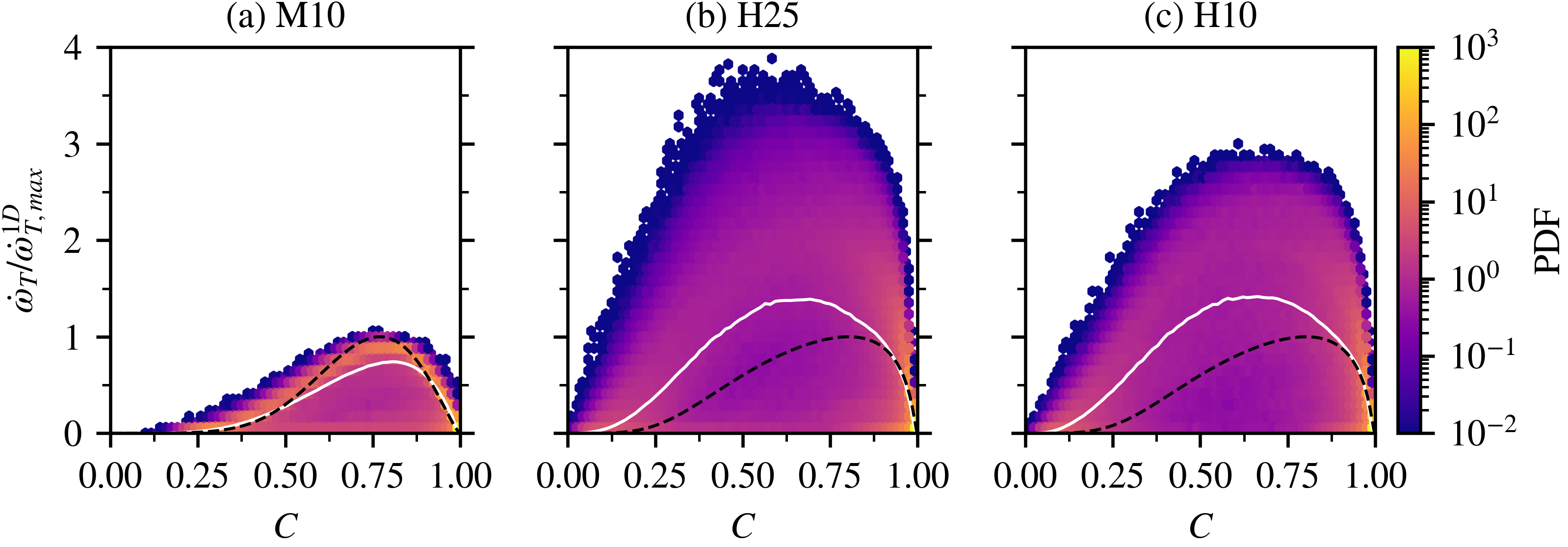}}
  \caption{Averages of the normalised heat release rate $\dot{\omega}_T/\dot{\omega}_{T,max}^{1D}$ conditioned by the progress variable $C$ (white lines) with the corresponding joint PDF (in logarithmic scale) at $t=4\tau$ for the three cases.
  The results for the corresponding 1-D unstretched flames (black dashed lines) are added for reference.}
  \label{fig:slot_jointpdf_hrr}
\end{figure}

The differences in the flame structures are examined in~\Cref{fig:slot_jointpdf_hrr}, which reports, for all cases, the mean normalised HRR $\dot{\omega}_T/\dot{\omega}_{T,\max}^{1D}$ conditioned by the progress variable $C$, together with the joint Probability Density Function (PDF) of these quantities and the values from the corresponding 1‑D unstretched laminar flames.
Ambient air dilution lowers the local equivalence ratio in the outer regions, producing low‑HRR zones across the full range of $C$ for all cases.
For the \ce{CH4} flame (see~\Cref{fig:slot_jointpdf_hrr}(a)), this leads to a conditional average lower than the 1-D unstretched flame and not coincident, as expected in a unity $Le$ flame.
Nevertheless, a narrow high-PDF region in correspondence of the 1-D unstretched laminar flame curve (black dashed line) can be identified.
Moreover, the peak value of HRR is located at $C=0.81$, close to the one of the 1-D case ({\it i.e.} $C=C^*=0.77$), and no relevant scatter is observed above the values of the 1-D flame (black dashed line).
On the other hand, for the \ce{H2} flames (see~\Cref{fig:slot_jointpdf_hrr}(b-c)), the variability of HRR induced by TD effects is observed, with a broader high-PDF region and a significant positive scatter above the mean (white solid line).
As a result, despite dilution by cold ambient air, the conditional means for the thermodiffusively unstable turbulent \ce{H2} flames remain above the corresponding unstretched 1‑D profile.
The deviation from the 1‑D flamelet is most pronounced in the preheat zone, as expected in the tThin Reaction Zone regime, shifting the HRR peak from $C=0.80$ in the laminar flame to $C^*=0.70$ in the turbulent cases.

\begin{figure}
  \centerline{\includegraphics[width=130mm]{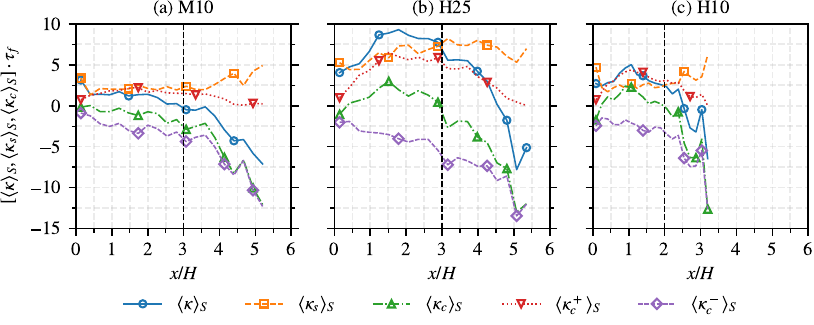}}
  \caption{Surface-averaged total stretch $\kappa$, strain component $\kappa_s$, curvature component $\kappa_c$ and its positive $\kappa_c^+$ and negative contributions $\kappa_c^-$, normalised by $\tau_f$, retrieved on the isosurface of $C=C^*$ at $t=4\tau$ for the three cases.}
  \label{fig:slot_strech_components_axis}
\end{figure}
\begin{figure}
  \centerline{\includegraphics[width=130mm]{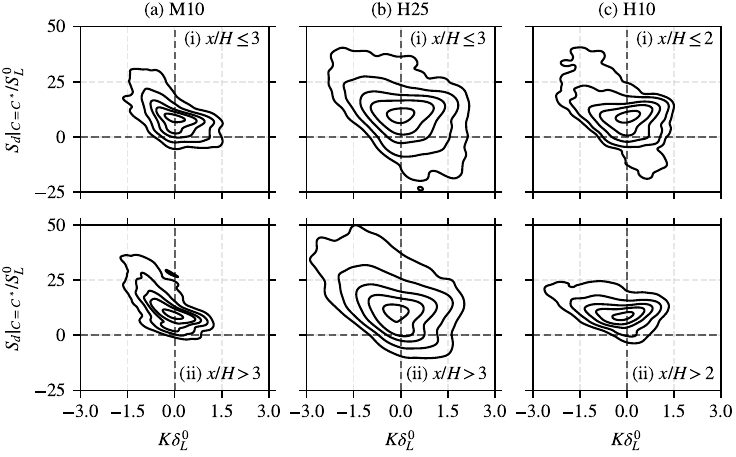}}
  \caption{Kernel Density Estimates of $\left.S_d\right|_{C=C^*}$ with respect to $K$ for the M10 (a), H25 (b), and H10 (c) cases in the downstream (i) and upstream (ii) flame regions, retrieved on the isosurface of $C=C^*$ at $t=4\tau$.}
  \label{fig:kde_stretch}
\end{figure}

As discussed in~\Cref{sec:intro,sec:modelling}, flame surface variations play a central role in the generation of combustion noise.
The differences in the local generation and destruction of the flame surface area among the different cases are thus characterised in~\Cref{fig:slot_strech_components_axis}, which reports the evolution along the stream-wise direction of the total stretch $\kappa$ and of its strain $\kappa_s$ and curvature $\kappa_c$ components, surface-averaged on the flame sheet defined by $C=C^*$ and normalised by the flame characteristic time $\tau_f=\delta_L^0/S_L^0$.
The different contribution of positive ($\kappa_c^+$) and negative ($\kappa_c^-$) curvature components is further accentuated by~\Cref{fig:kde_stretch}, which reports the Kernel Density Estimate (KDE) between $\left.S_d\right|_{C=C^*}$ and the (geometrical) flame curvature $K$ for the three cases in the upstream and downstream parts of the flame.
Further details on the definition of the stretch components and of the surface average operator are provided in Appendix~\ref{app:stretch}.

Focusing on the M10 case first (see~\Cref{fig:slot_strech_components_axis}(a)), the upstream region of the flame ($x/H\leq3$) shows a strong balance between positive ($\langle\kappa_c^+\rangle_S$) and negative ($\langle\kappa_c^-\rangle_S$) curvature components, with the two assuming almost the same value.
This results in a slightly positive total stretch and indicates a limited enhancement of flame surface controlled by strain.
This is further highlighted by the KDE in~\Cref{fig:kde_stretch}(a-i), which shows a balance between positive and negative values of $K$.
In the downstream region ($x/H>3$), however, the contribution of negative curvature becomes dominant, due to the presence of extreme cusp regions~\citep{han2008roles,coulon2023direct}, leading to the detachment of flame pockets at the tip.
The dominance of negative curvature in this region is further highlighted by the KDE in~\Cref{fig:kde_stretch}(a-ii), which shows an unbalance towards $K<0$.
A stronger enhancement of $\left.S_d\right|_{C=C^*}$ is present as $K$ becomes further negative, due to the so-called kinematic restoration~\citep{peters1999turbulent}, {\it i.e.} the tendency of negatively curved flame elements to accelerate in order to smooth out wrinkles and reduce flame surface area.
This leads to a strong local surface destruction induced by flame--flame interactions~\citep{dave2020evolution}, which has been associated to high-frequency noise generation in hydrocarbon flames~\citep{brouzet2019annihilation}.

Moving to the H25 case (see~\Cref{fig:slot_strech_components_axis}(b)), a stark difference appears in the first region of the flame ($x/H\leq 3$), with an unbalance of the curvature term $\langle\kappa_c\rangle_S$ toward the positive contribution, which adds up to the turbulence-related strain $\langle\kappa_s\rangle_S$.
This results in a strongly positive total stretch $\langle\kappa\rangle_S$ and, consequently, in an enhancement of the flame surface area.
This is caused by TD instabilities, which lead to a pronounced wrinkling of the flame surface and to the strong enhancement of the local HRR in the highly curved regions (see~\Cref{fig:slot_isosurf}(b)). The differences with respect to the M10 case (see~\Cref{fig:kde_stretch}(a-i)) are further highlighted from the KDE in~\Cref{fig:kde_stretch}(b-i), where a wider distribution is observed, especially in the first quadrant, in which both $\left.S_d\right|_{C=C^*}$ and $K$ are positive and increase together, thereby promoting flame surface generation.
In the rear part of the jet ($x/H>3$), a decrease of the curvature term $\langle\kappa_c\rangle_S$ towards negative values is observed (see~\Cref{fig:slot_strech_components_axis}(b)), promoting surface destruction at the flame tip.
Differently from the M10 case (see~\Cref{fig:slot_strech_components_axis}(a)), however, the switch from flame surface generation to destruction, {\it i.e.} where $\langle\kappa\rangle=0$, occurs further downstream (see~\Cref{fig:slot_strech_components_axis}(b)).
In other terms, the contribution of positive curvature remains relevant also in the downstream part of the jet, as highlighted from the KDE in~\Cref{fig:kde_stretch}(b-ii).
This attenuates the impact of flame destruction on the surface variation at the tip, marking a substantial difference with respect to the \ce{CH4} flame.

Similar features can be observed also in the H10 case (see~\Cref{fig:slot_strech_components_axis}(c)), with the main difference being the reduction of the strain term $\langle\kappa_s\rangle_S$, especially in the first region (\mbox{$x/H\leq 2$}), due to the reduced bulk velocity $U_B$.
The comparison of the behaviours at the flame tip of the M10 (see~\Cref{fig:kde_stretch}(a-ii)) and H10 (see~\Cref{fig:kde_stretch}(c-ii)) cases, which share the same value of $U_B$, allows to further highlight the influence of TD effects.
Indeed, a much more limited increase of $\left.S_d\right|_{C=C^*}$ is observed for $K<0$ in the \ce{H2} flame, a result that was related by~\citet{coulon2023direct} to the shift of the diffusive contribution of $\left.S_d\right|_{C=C^*}$ to positive values.
Consequently, the values reached for positive and negative curvatures are comparable in the H10 case, confirming the reduced importance of annihilation phenomena on the flame surface variation at the tip.

\subsection{Acoustic radiation}
\label{sec:acoustic}

\begin{figure}
  \centerline{\includegraphics[width=130mm]{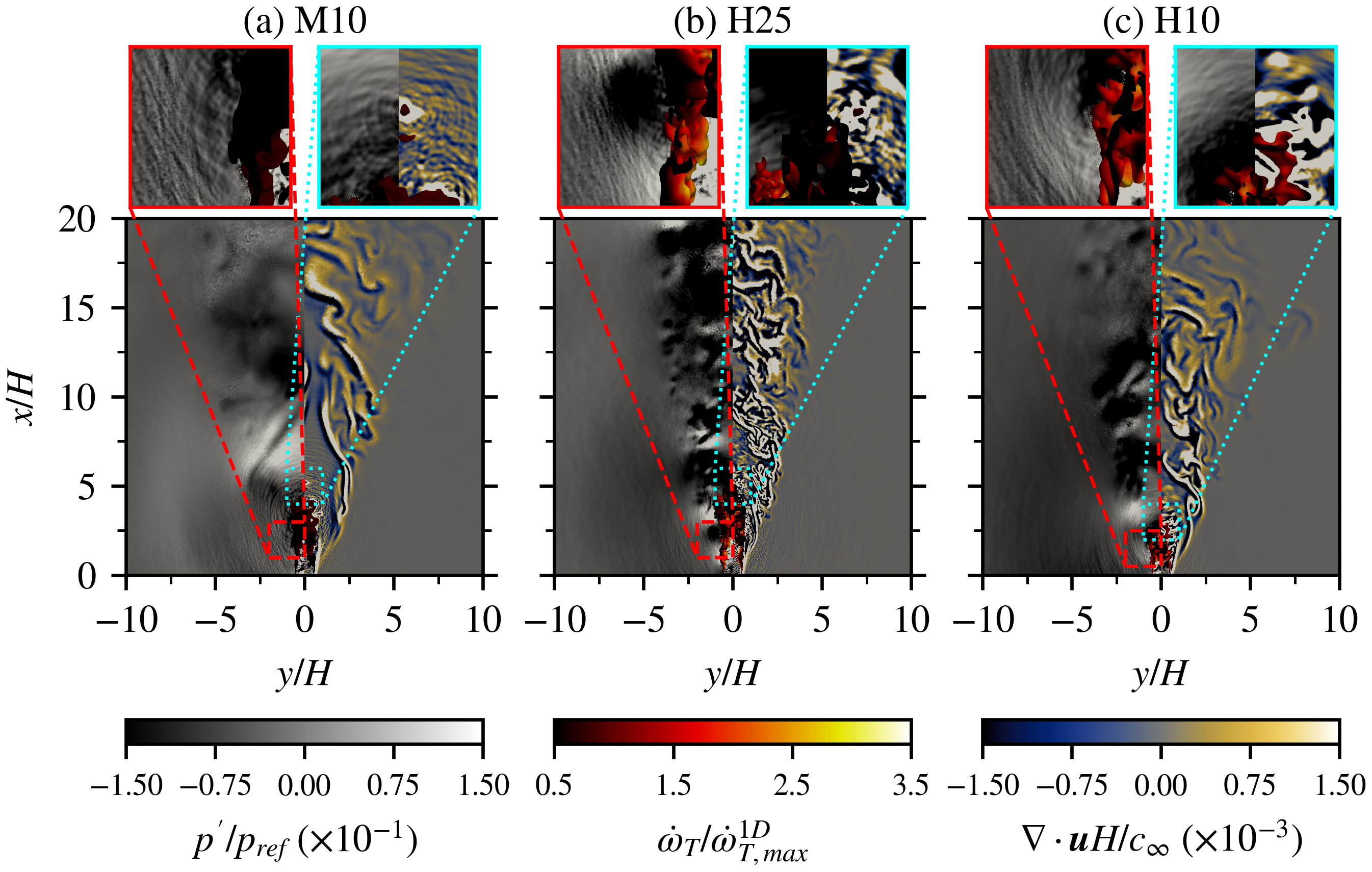}}
  \caption{Normalised fields of fluctuating pressure $p'/p_{ref}$ (left) and dilatation $\nabla\cdot\boldsymbol{u}H/c_\infty$ (right), and isosurfaces of $C=C^*$ coloured by the normalised heat release rate $\dot{\omega}_T/\dot{\omega}_{T,max}^{1D}$ at $t=4\tau$ in the $xy$ plane for the three cases at $t=4\tau$.}
  \label{fig:slot_pressure_dilatation_xy}
\end{figure}

A qualitative assessment of the acoustic field is provided in~\Cref{fig:slot_pressure_dilatation_xy}, which reports instantaneous fields of the normalised fluctuating pressure $p'=p-\overline{p}$ (on the left) and dilatation $\nabla\cdot\boldsymbol{u}$ (on the right) for the three cases in the $xy$ plane.
The dilatation is the divergence of the flow velocity, and is widely used in aeroacoustic studies to identify sound~\citep{colonius1997sound,talei2013direct,talei2014comparative}, being it related to the far field pressure~\citep{colonius1997sound}:
\begin{equation}
    \frac{\partial p}{\partial t} + \rho_\infty c_\infty^2\nabla\cdot\boldsymbol{u} = 0.
\end{equation}
The red dashed insets in~\Cref{fig:slot_pressure_dilatation_xy} highlight the acoustic radiation from the centre of the flame brushes, while the cyan dotted insets are taken at the flame tip.
The isosurfaces of $C=C^*$ coloured by the local normalised HRR, previously reported in~\Cref{fig:slot_isosurf}, are added for reference.

By comparing the \ce{CH4} flame (see~\Cref{fig:slot_pressure_dilatation_xy}(a)) with the \ce{H2} ones (see~\Cref{fig:slot_pressure_dilatation_xy}(b-c)), some first qualitative differences can be observed in the acoustic radiation.
Indeed, the \ce{CH4} flame shows a strong contribution originating at the flame tip (see close‑up in the cyan box), due to the rapid consumption of the detached flame pockets, a well-known noise generation mechanism for these flames (see~\Cref{sec:intro}).
For the \ce{H2} flames, on the other hand, the contribution from the flame tip seems less evident, consistently with the differences in the stretch components observed in this region in~\Cref{sec:flame_turbulence_int}.
Moreover, relevant differences are present also in the outer shear layer, as highlighted by the dilatation field (see close‑up in the red box).
Indeed, all flames present K-H shear layer instability structures, which arise from the velocity gradient between the high-velocity jet and the low-velocity ambient air and are a known source of acoustic radiation in turbulent jets~\citep{jordan2013wave}.
Nevertheless, the M10 case shows only larger scale phenomena compared to the \ce{H2} flames. While for the H25 case this difference could be imputed to the higher bulk velocity, the presence of similar features in the low-velocity H10 case suggests that the observed finer scale structures are related to the different transport properties of \ce{CH4} and \ce{H2}.

\begin{figure}
  \centerline{\includegraphics[width=130mm]{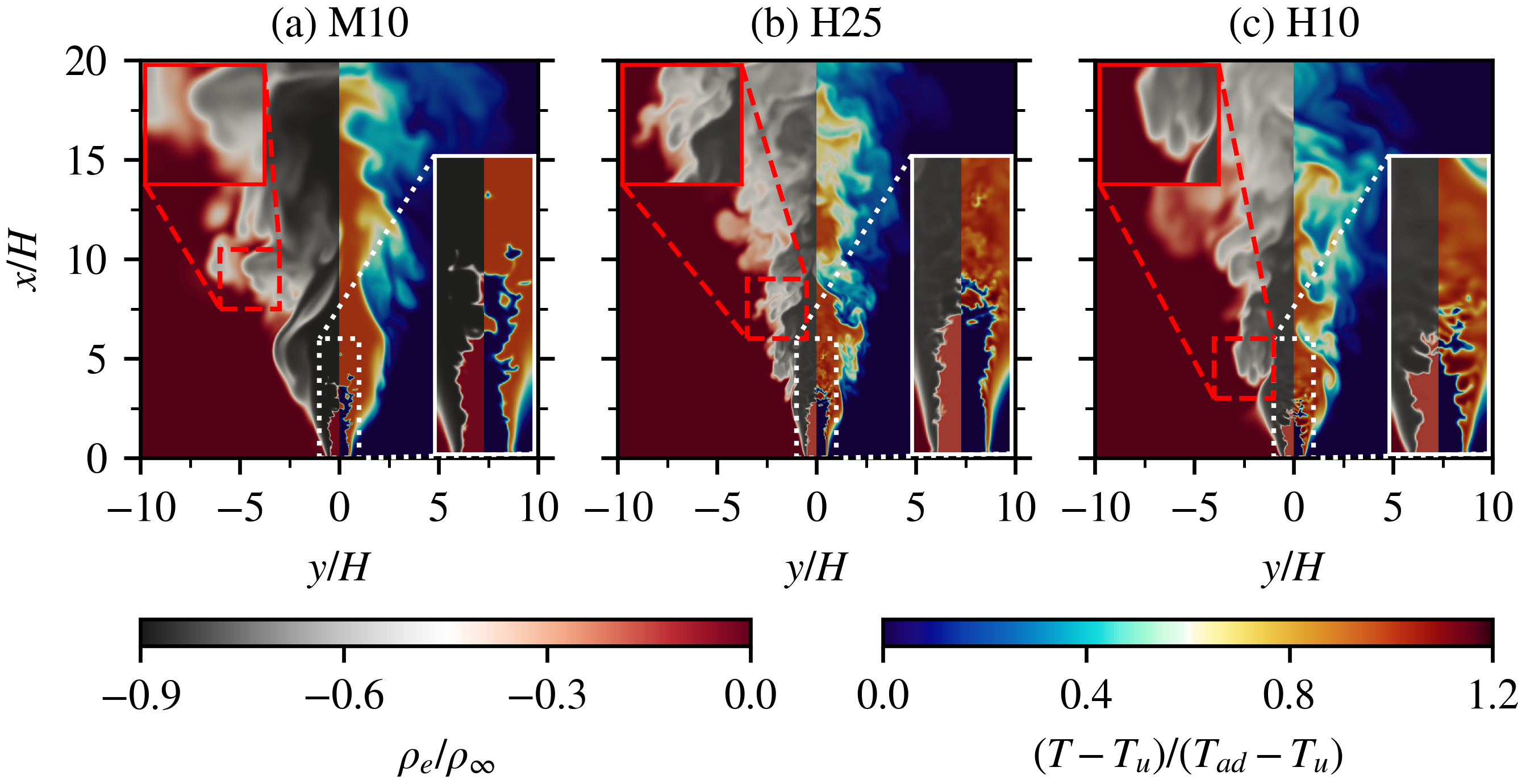}}
  \caption{Normalised fields of excess density $\rho_e/\rho_\infty$ (left) and temperature $(T-T_u)/(T_{ad}-T_u)$ (right) in the $xy$ plane for the three cases at $t=4\tau$.}
  \label{fig:slot_temp_rhoe_xy}
\end{figure}

To further investigate the differences in the outer shear layer,~\Cref{fig:slot_temp_rhoe_xy} reports the normalised fields of excess density $\rho_e$ (on the left) and temperature (on the right) in the $xy$ plane for the three cases.
The excess density is defined as:
\begin{equation}
    \rho_e = \rho - \rho_\infty - \frac{p-p_\infty}{c_\infty^2},
\end{equation}
and is non-zero where the entropy is strongly different from the surroundings~\citep{dowling2015combustion}.
Its spatial and temporal gradient is present in the acoustic analogy for reacting flows by~\citet{dowling1992thermoacoustic} as indirect combustion noise source, being it associated with entropy fluctuations~\citep{dowling2015combustion}.

Some discrepancies can be observed from the temperature fields in the burnt gases region.
The \ce{CH4} flame presents a more uniform distribution, with temperature equal to the adiabatic value, while the \ce{H2} flames show strong fluctuations, even reaching super-adiabatic conditions, an indicator of TD effects~\citep{giannakopoulos2015consistent,berger2024effects}, as highlighted in the white inset in the bottom right corner.
The same TD mechanism, based on the \ce{H2} non-equidiffusion, appears to impact also the outer shear layer between cold ambient air and hot combustion products, which exhibits a more corrugated interface in the \ce{H2} flames compared to the \ce{CH4} one.
As a result, the mixing of the two fluids is promoted, reducing the radial extension of the burnt gases region.
Moreover, finer scale structures, propagating toward the cold ambient air, are present in the \ce{H2} cases, while the \ce{CH4} flame presents only large scale structures, as observable from the field of $\rho_e$ in the red inset in the top left corner.
This observation suggests that \ce{H2} non-equidiffusion may contribute to modifying the shear layer dynamics at this non-reacting interface, possibly through its effect on the density field variability in the burnt gases region.
The influence of these differences on the acoustic field will be further discussed in~\Cref{sec:spod_noise}, by analysing the coherent structures resulting from the SPOD of pressure in the different cases.

\begin{figure}
  \centerline{\includegraphics[width=130mm]{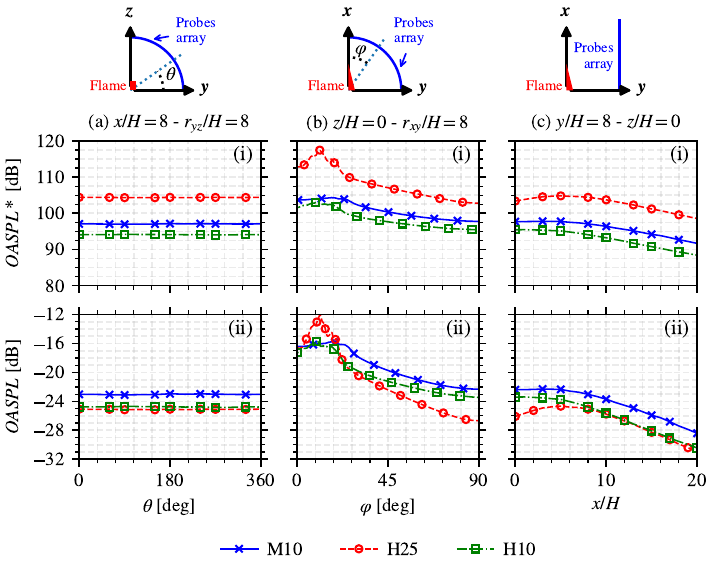}}
  \caption{Non-normalised $OASPL^*$ (i) and normalised $OASPL$ (ii) for the three cases as a function of the azimuthal angle $\theta=\tan^{-1}(z/y)$ in the $yz$ plane at a radial distance $r_{yz}=(y^2+z^2)^{1/2}$ equal to $8H$ (a), as a function of the polar angle $\varphi=\tan^{-1}(y/x)$ in the $xy$ plane at a radial distance $r_{xy}=(x^2+y^2)^{1/2}$ equal to $8H$ (b), and as a function of the streamwise position $x/H$ for fixed coordinates $y=8H$ and $z=0$ (c). The schemes on top show the locations of the probes with respect to the flame.}
  \label{fig:directivity_no_norm}
\end{figure}

The directivity of the acoustic radiation is investigated in~\Cref{fig:directivity_no_norm}.
First, the non-normalised Overall Sound Pressure Level ($OASPL^*$) is considered (see~\Cref{fig:directivity_no_norm}(i)), defined as:

\begin{equation}
    OASPL^*=20\log_{10}\left(\frac{p'_{rms}}{p_{0}}\right),
\end{equation}
where $p'_{rms}$ is the temporal r.m.s. of the local pressure fluctuation, and $p_0=2.0\times 10^{-5}$~Pa is a reference acoustic pressure~\citep{candel2009flame}.
\Cref{fig:directivity_no_norm} reports the $OASPL^*$ as a function of the azimuthal angle $\theta$ in the $yz$ plane at a radial distance $r_{yz}=(y^2+z^2)^{1/2}$ equal to $8H$ (a), as a function of the polar angle $\varphi$ in the $xy$ plane at a radial distance $r_{xy}=(x^2+y^2)^{1/2}$ equal to $8H$ (b), and as a function of the stream-wise position $x/H$ for fixed coordinates $y=8H$ and $z=0$ (c).
On top of each plot, a scheme is reported to show the position of the probes with respect to the flame.

A nearly-isotropic behaviour is observed in the $yz$ plane (see~\Cref{fig:directivity_no_norm}(a-i)), coherently with the monopolar nature of direct combustion noise~\citep{truffaut1998etude}.
Moving to the directivity in the $xy$ plane (see~\Cref{fig:directivity_no_norm}(b-i)), the region impacted by the hydrodynamic and indirect noise sources is evident, with an enhancement of the $OASPL^*$ observable in correspondence of the outer shear layer, {\it i.e.} $\varphi=15$° for the \ce{H2} flames and $\varphi=30$° for the \ce{CH4} one, coherently with the differences in the radial extension of the burned gases region shown in~\Cref{fig:slot_temp_rhoe_xy}.
Finally, a directivity is observed far from the jet (see~\Cref{fig:directivity_no_norm}(c-i)) for $x/H < 8$, where the acoustic field may be impacted by the coflow inlet boundary~\citep{brouzet2021impact}.
For $x/H \geq 8$, instead, the $OASPL^*$ varies linearly with the distance, coherently with~\Cref{eq:noise_hrr}, suggesting that in this region the acoustic radiation is related mostly, if not solely, to direct combustion noise radiated from the flame.

In all cases shown in~\Cref{fig:directivity_no_norm}(i), it is observed that the intensity of the acoustic radiation is stronger for the H25 flame, followed by the M10 and by the H10 ones.
This is consistent with the higher thermal power of the H25 case with respect to the other ones (see~\Cref{tab:flow_params}).
To remove this bias,~\Cref{fig:directivity_no_norm}(ii) reports the same directivity analysis of~\Cref{fig:directivity_no_norm}, but considering the normalised Overall Sound Pressure Level:
\begin{equation}
    OASPL=20\log_{10}\left(\frac{p'_{rms}}{p_{ref}}\right),
\end{equation}
where $p_{ref}$ is the reference pressure previously defined in~\Cref{eq:p_ref}.
Evidently, the directivity of each flame is not impacted by the normalisation, since this induces, in logarithmic scale, only a vertical shift of the values.
Nevertheless, stark differences appear in the comparison of the different flames at the same spatial location.
In the $yz$ plane (see~\Cref{fig:directivity_no_norm}(a-ii)), where flame-generated noise prevails, the two \ce{H2} flames have almost identical $OASPL$ values.
Moving to the directivity in the $xy$ plane (see~\Cref{fig:directivity_no_norm}(b-ii)), the H25 flame, characterised by a higher velocity, dominates in the near field, while the slower M10 and H10 flames present a stronger $OASPL$ for $\varphi\geq 30°$.
Finally, a superposition of the $OASPL$ values is observed for the two \ce{H2} flames far from the jet for \mbox{$x/H\geq 7$} (see~\Cref{fig:directivity_no_norm}(c-ii)).
Overall, even though a collapse of all cases is not achieved, probably due to the fact that the simplified approach adopted in the definition of $p_{ref}$ (see Appendix~\ref{app:pressure}) does not include any correlation volume effect~\citep{rajaram2006premixed}, the normalisation with $p_{ref}$ still allows for a more adequate comparison of the different cases, sustaining the validity of this choice.

\begin{figure}
  \centerline{\includegraphics[width=70mm]{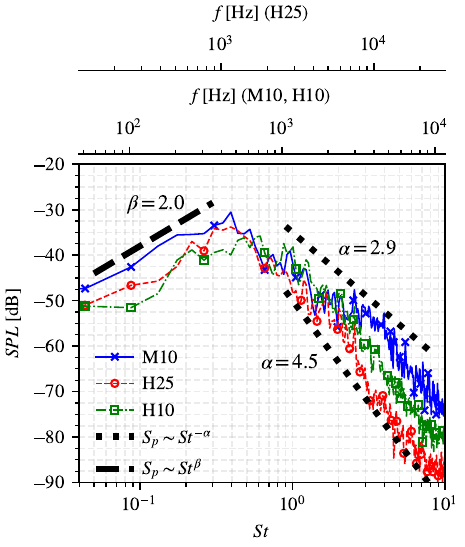}}
  \caption{$SPL$ spectra at $x/H=8$ and $r_{yz}/H=8$, averaged over 12 equally distributed probes in the azimuthal direction, for the three cases, as a function of the Strouhal number $St$ (bottom) and of the frequency $f$ (top).}
  \label{fig:pressure_spectra_farfield}
\end{figure}

Based on the analysis of directivity in~\Cref{fig:directivity_no_norm}, the acoustic radiation of the three flames is compared hereafter considering the pressure traces at $x/H=8$ and $r_{yz}/H=8$.
Given the monopolar nature of the acoustic radiation in this region (see~\Cref{fig:directivity_no_norm}(b)), the average of the pressure traces of 12 equally distributed probes in the azimuthal direction is considered, thus exploiting ergodicity to improve the quality of the spectral analysis.
The Sound Pressure Level ($SPL$) spectra for the three cases are reported in~\Cref{fig:pressure_spectra_farfield}, with the $SPL$ defined as:
\begin{equation}
    SPL=10\log_{10}\left(\frac{S_p\Delta f}{p_{ref}^2}\right),
\end{equation}
where $\Delta f$ is the frequency resolution.
The resulting spectra are coherent with the literature on combustion noise~\citep{rajaram2009acoustic,liu2015modelling}, with power law dependencies in the low- (black dashed line) and high-frequency (black dotted lines) regions.
The M10 and H25 cases both peak at $St\approx 0.4$, where $St=fH/U_B$ is the Strouhal number, while the H10 case peaks at $St\approx0.6$.
This shift in the peak $St$ value is to be expected, since the peak frequency of the combustion noise spectrum scales as $U_B/L_f$~\citep{rajaram2009acoustic}, meaning that the M10 and H25 cases, with comparable flame brush length, have similar peak $St$ values, while the shorter H10 flame peaks at higher $St$.
In terms of frequency, however, the \ce{H2} flames are both shifted to higher values with respect to the \ce{CH4} one, as highlighted from the values of $f$ (in Hz) reported in the upper part of~\Cref{fig:pressure_spectra_farfield}.

Similar features are observed in the low-frequency range among the three cases, with the same slope $\beta=2.0$, a value in agreement with previous works in the literature~\citep{rajaram2009acoustic,liu2015modelling,kumar2025spectral}.
On the other hand, more enhanced discrepancies are observed in the high-frequency range, above the peak $St$ value.
Indeed, the M10 case shows a slope (in logarithmic scale) $\alpha=2.9$, coherent with the ranges $2.2\leq\alpha\leq3.4$ found by~\citet{belliard1997etude} and $2.1\leq\alpha\leq3.2$ found by~\citet{rajaram2009acoustic} for premixed hydrocarbon unconfined turbulent flames, while a steeper decay is observed for both \ce{H2} flames, with $\beta=4.5$.
Still, for $St<1$, the (normalised) spectra of the three flames are almost superimposed, while the steeper decay of the \ce{H2} flames is more accentuated for $St>1$, where, conversely, an enhancement of the $SPL$ values is observed for the M10 case.
This is significant, since the region at $St>1$ was identified by~\citet{brouzet2019annihilation} as the one where flame annihilation events are the dominant noise sources for premixed hydrocarbon flames, being these sudden, hence high-frequency phenomena.
Therefore, the reduced acoustic radiation in this high-$St$ band for the \ce{H2} flames is coherent with a reduced importance of flame annihilation as noise source, due to its slower dynamics~\citep{talei2012parametric,talei2013direct}.
This is consistent with the analysis of the flame--turbulence interaction in~\Cref{sec:flame_turbulence_int} and will be detailed in~\Cref{sec:td_effects_noise}.

\begin{figure}
  \centerline{\includegraphics[width=130 mm]{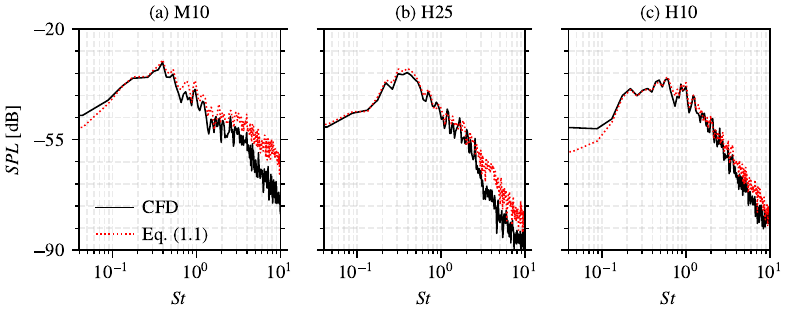}}
  \caption{Normalised $SPL$ spectra at $x/H=8$ and $r_{yz}/H=8$ directly extracted from the CFD results and averaged over 12 equally distributed probes in the azimuthal direction (solid black lines) and computed from $\text{d}q/\text{d}t$ via \Cref{eq:noise_hrr} (red dotted lines) for the three cases.}
  \label{fig:pressure_hrr_spectra}
\end{figure}

To further explain the differences observed between the two fuels, the acoustic radiation is related to the time derivative of the volume integrated HRR $\text{d}q/\text{d}t$ by applying \Cref{eq:noise_hrr} \textit{a posteriori}.
The distance $\left|\boldsymbol{x}\right|$ in~\Cref{eq:noise_hrr} is here taken by considering the origin in the mean centre of the flame, {\it i.e.} from the point on the $x$-axis at $x=L_f/2$.
\Cref{fig:pressure_hrr_spectra} compares the $SPL$ spectra directly recorded during the simulation (black solid lines) and reconstructed from the HRR signal (red dotted lines) for the three cases.
An almost perfect match is found between the recorded and reconstructed $SPL$ for $St\leq 3$, confirming that the unsteady temporal fluctuation of HRR is the dominant noise source in this region far from the reacting jet.
Some minor discrepancies are found at higher $St$ ($St>3$), especially for the M10 case (see~\Cref{fig:pressure_hrr_spectra}(a)).

\section{Impact of thermodiffusive effects on combustion noise}
\label{sec:td_effects_noise}

In this Section, we intend to relate the differences in the flame--turbulence interaction observed among the three cases in~\Cref{sec:flame_turbulence_int} to the discrepancies in the acoustic features found in~\Cref{sec:acoustic}.
The main focus is on the impact of TD effects on the flame surface generation and destruction processes (see~\Cref{sec:flame_surface}) and on the local HRR variations (see~\Cref{sec:hrr_noise}), to gain a further understanding of the underlying mechanisms responsible for combustion noise generation in lean premixed \ce{H2} flames.

\subsection{Flame surface area variations and stretch effects}
\label{sec:flame_surface}

\begin{figure}
  \centerline{\includegraphics[height=50mm]{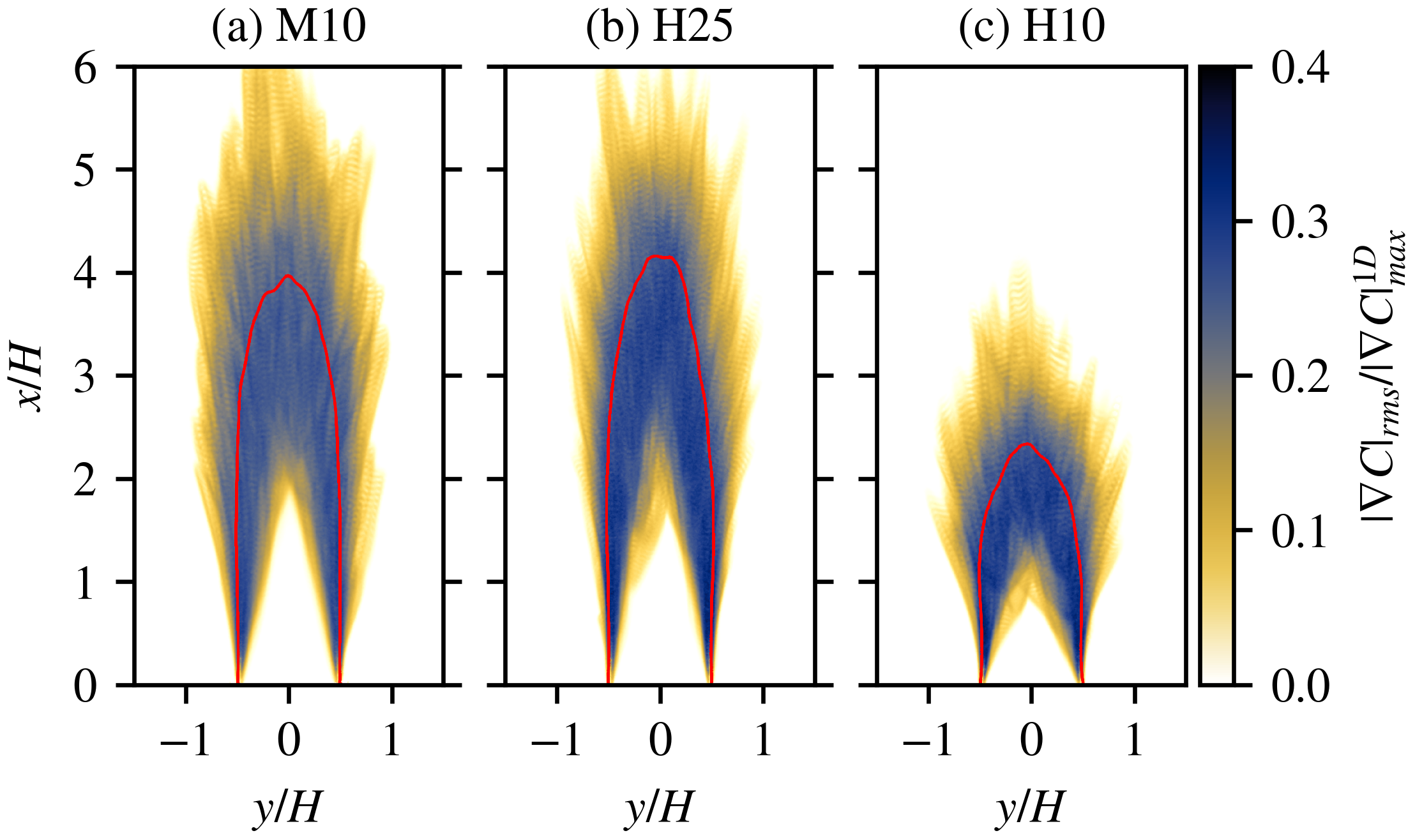}}
  \caption{Normalised temporal r.m.s. fields of the magnitude of the gradient of progress variable $\left|\nabla C\right|_{rms}/\left|\nabla C\right|_{max}^{1D}$ in the $xy$ plane for the three cases.
  The red lines denote contours of mean progress variable $\widetilde{C}=C^*$.}
  \label{fig:gradc_rms}
\end{figure}

\Cref{fig:gradc_rms} shows the spatial field in the $xy$ plane of the local temporal r.m.s. of the magnitude of the gradient of progress variable $\left|\nabla C\right|_{rms}$, which is related to the flame surface area fluctuations (see \Cref{eq:turb_flame_surf}), normalised by its maximum value in the corresponding 1-D laminar flame.
The isocontour at $\widetilde{C}=C^*$ is reported in red to denote the mean flame position.
Overall, the distributions are comparable among the different flames in terms of intensity of the fluctuations.
This is consistent with the similarities in the temporal traces of $A_T/A_0$ reported in~\Cref{fig:st_at_i0}(b).
Nevertheless, two qualitative differences can be observed.
First, the \ce{CH4} flames presents more widespread fluctuations in the downstream region (i.e., for $\widetilde{C}>C^*$), coherently with the analysis of stretch components in~\Cref{sec:flame_turbulence_int} and, in particular, with the detachment and subsequent rapid consumption of flame pockets at the tip.
Secondly, the \ce{H2} flames show more intense fluctuations of $\left|\nabla C\right|$ on the fresh gases side, i.e., for $\widetilde{C}<C^*$, which well relate to the generation of flame surface area due to finger-like structures propagating towards the fresh gases, which are associated with TD instabilities~\citep{berger2022synergistic,rieth2022enhanced,coulon2023direct}.

\begin{figure}
  \centerline{\includegraphics[width=130 mm]{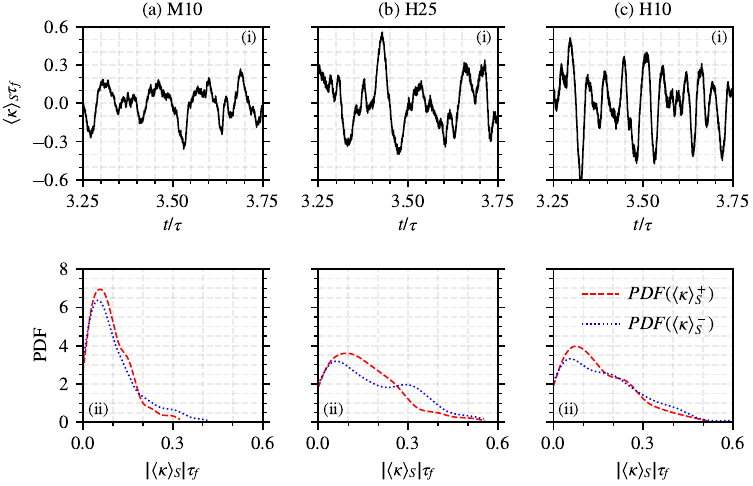}}
  \caption{Normalised global flame stretch traces $\langle\kappa\rangle_S\tau_f$ (i) and conditional PFDs of its positive $\langle\kappa\rangle_S^+$ and negative $\langle\kappa\rangle_S^-$ contributions (ii) for the three cases.}
  \label{fig:globa_stretch_pdf}
\end{figure}

In spite of the apparent similarities, the impact of stretch effects on the temporal variation of the flame surface area in the thermodiffusively unstable \ce{H2} flames is non-negligible.
In this sense,~\Cref{fig:globa_stretch_pdf} shows the temporal evolution of the total surface-averaged flame stretch $\langle\kappa\rangle_S$ (top), related to the temporal derivative of $A_T$ (see Appendix~\ref{app:stretch}), together with the corresponding PDFs of its positive ($\langle\kappa\rangle_S^+$) and negative ($\langle\kappa\rangle_S^-$) contributions (bottom), i.e., the PDFs of $\langle\kappa\rangle_S$ conditioned by its sign.
From the temporal traces (see~\Cref{fig:globa_stretch_pdf}(i)), it can be observed that the \ce{H2} flames are prone to larger values of stretch, hence of flame surface deformation.
This is further highlighted by the PDF distributions (see~\Cref{fig:globa_stretch_pdf}(ii)), showing that the \ce{H2} flames reach higher values of mean flame stretch, in both the positive and negative directions.

Moreover, the strengths of the positive and negative contributions differ when passing from \ce{CH4} to \ce{H2}.
Indeed, while all flames are statistically stationary, meaning that the positive and negative contributions balance out on average~\citep{brouzet2021impact}, an increased importance of the positive term (i.e., of the flame surface generation component) in the statistical distribution is observed when passing from the thermodiffusively stable \ce{CH4} flame to the unstable \ce{H2} ones.
This can be quantified by the skewness of the PDF distribution, which passes from -0.37 for the M10 case to -0.16 and -0.25 for the H25 and H10 cases, respectively.
The stronger negative skewness for the M10 flame is observable in~\Cref{fig:globa_stretch_pdf}(a-ii) from the longer tail of the PDF of $\langle\kappa\rangle_S^-$ (blue dotted curve) with respect to $\langle\kappa\rangle_S^+$ (red dashed curve).
This indicates that extreme negative stretch events, albeit rare, are not balanced by comparable positive stretch ones.
Physically, this means that the \ce{CH4} flame is more prone to strong destructive flame annihilation events, which play a major role in the generation of combustion noise at high frequencies.
On the other hand, the more positive skewness for the \ce{H2} flames, indicating a reduced asymmetry toward the negative side, sustains the increased importance of flame generation previously discussed in~\Cref{sec:flame_turbulence_int}.
As a result, the importance of flame annihilation on the overall flame dynamics and, consequently, on the high-frequency combustion noise, is reduced.

\begin{figure}[t!]
  \centerline{\includegraphics[width=130 mm]{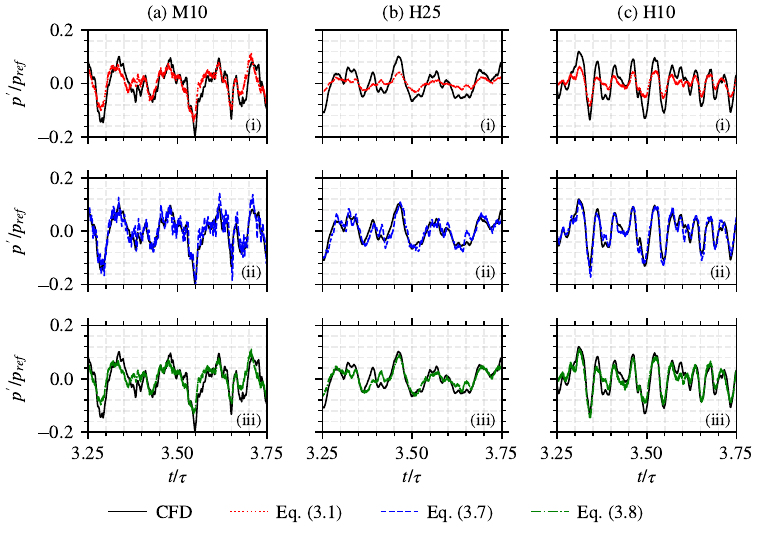}}
  \caption{Normalised pressure fluctuation traces $p'/p_{ref}$ directly extracted  from the CFD results and averaged over 12 equally distributed probes in the azimuthal direction (solid black lines) and computed (i) from $\text{d}A/\text{d}t$ with~\Cref{eq:p_area_le1} (red dotted lines), (ii) from $\text{d}(I_0A)/\text{d}t$ with~\Cref{eq:p_area_i0} (blue dashed lines) and (iii) from $\overline{I_0}\text{d}A/\text{d}t$ with~\Cref{eq:p_area_i0_avg} (green dash-dotted lines) for the three cases.}
  \label{fig:pressure_area}
\end{figure}

To further relate the radiated sound to the flame surface dynamics, the theoretical framework developed in~\Cref{sec:modelling} is here applied to the DNS dataset.
\Cref{fig:pressure_area} compares the temporal traces of normalised fluctuating pressure $p'/p_{ref}$, directly extracted from the calculations at the same location of the spectra of~\Cref{fig:pressure_spectra_farfield}, with those reconstructed {\it a posteriori} from $\text{d}A_T/\text{d}t$ using the classical flamelet theory~\citep{abugov1978acoustic,clavin1991turbulent} in \Cref{eq:p_area_le1} (see~\Cref{fig:pressure_area}(i)) or the modified version developed in the present work to include stretch effects, either in the temporal derivative (see~\Cref{fig:pressure_area}(ii)) by applying \Cref{eq:p_area_i0} or in the mean (see~\Cref{fig:pressure_area}(iii)) by using \Cref{eq:p_area_i0_avg}.
While the amplitude of $p'/p_{ref}$ is well predicted by \Cref{eq:p_area_le1} for the M10 case, an underestimation is consistently found for the \ce{H2} flames, confirming the failure of the classical combustion noise theory for these flames.
This is quantified by the concordance correlation coefficient $r_c$~\citep{lawrence1989concordance}, which takes into account both the linear association and the local distance between the values of the two distributions.
Its value is equal to 0.83 for the M10 case, while it drops to 0.55 and 0.68 for the H25 and H10 cases, respectively.
By introducing the stretch factor $I_0$ in the temporal derivative using \Cref{eq:p_area_i0} (see~\Cref{fig:pressure_area}(ii)), no relevant differences, as expected, are observed for the \ce{CH4} flame, since this is not affected by stretch effects, {\it i.e.} $I_0\approx 1$ (see~\Cref{fig:st_at_i0}(c)).
However, a much better agreement is observed for the \ce{H2} flames, with the recorded and reconstructed $p'$ signals almost superimposed, and the values of $r_c$ being not lower than 0.9 (a perfect match corresponds to $r_c=1$).
The validity of the proposed extension of the flamelet theory for thermodiffusively unstable flames is thus confirmed.
Most notably, the use of the time-averaged stretch factor $\overline{I_0}$ (see~\Cref{fig:pressure_area}(iii)) has only a limited negative influence on the amplitude of the predicted noise radiation.
Therefore, \Cref{eq:p_area_i0_avg} represents a valid solution to assess the radiated sound from the flame surface area of thermodiffusively unstable lean premixed \ce{H2} flames, especially when direct measurements of pressure or HRR are not feasible.
It is worth noticing that $\overline{S_T}/S_L^0>\overline{I_0}$ (see~\Cref{fig:st_at_i0}(b-c)), which may explain the over-prediction of the noise level observed by~\citet{talei2013direct} when applying \Cref{eq:p_area_le1} with $S_L^0$ replaced by the time-averaged consumption speed.

From a physical point of view, the validity of~\Cref{eq:p_area_i0,eq:p_area_i0_avg} means that stretch effects, by promoting the wrinkling of the surface of thermodiffusively unstable flames, effectively enhance the generation of sound.
Nevertheless, \Cref{eq:p_area_i0_avg} also highlights that the amplitude of the radiated sound is modulated not only by $\overline{I_0}$, but also by the laminar flame speed $S_L^0$ and by the expansion ratio across the flame front $\rho_u/\rho_b$.
In the present case, while the surface fluctuations are comparable among the \ce{CH4} and \ce{H2} flames, given that they are subject to analogous turbulence fields, and $S_L^0$ is the same by choice (see~\Cref{tab:lam_properties}), the lean \ce{H2} flames feature a lower value of the expansion ratio $\rho_u/\rho_b$ with respect to the stoichiometric \ce{CH4} one, given the differences in the values of the adiabatic flame temperature $T_{ad}$ (see~\Cref{tab:lam_properties}).
This counterbalances the enhancement of combustion noise generated by TD effects through the stretch factor.

\subsection{Local heat release rate dynamics}
\label{sec:hrr_noise}

\begin{figure}
  \centerline{\includegraphics[height=50mm]{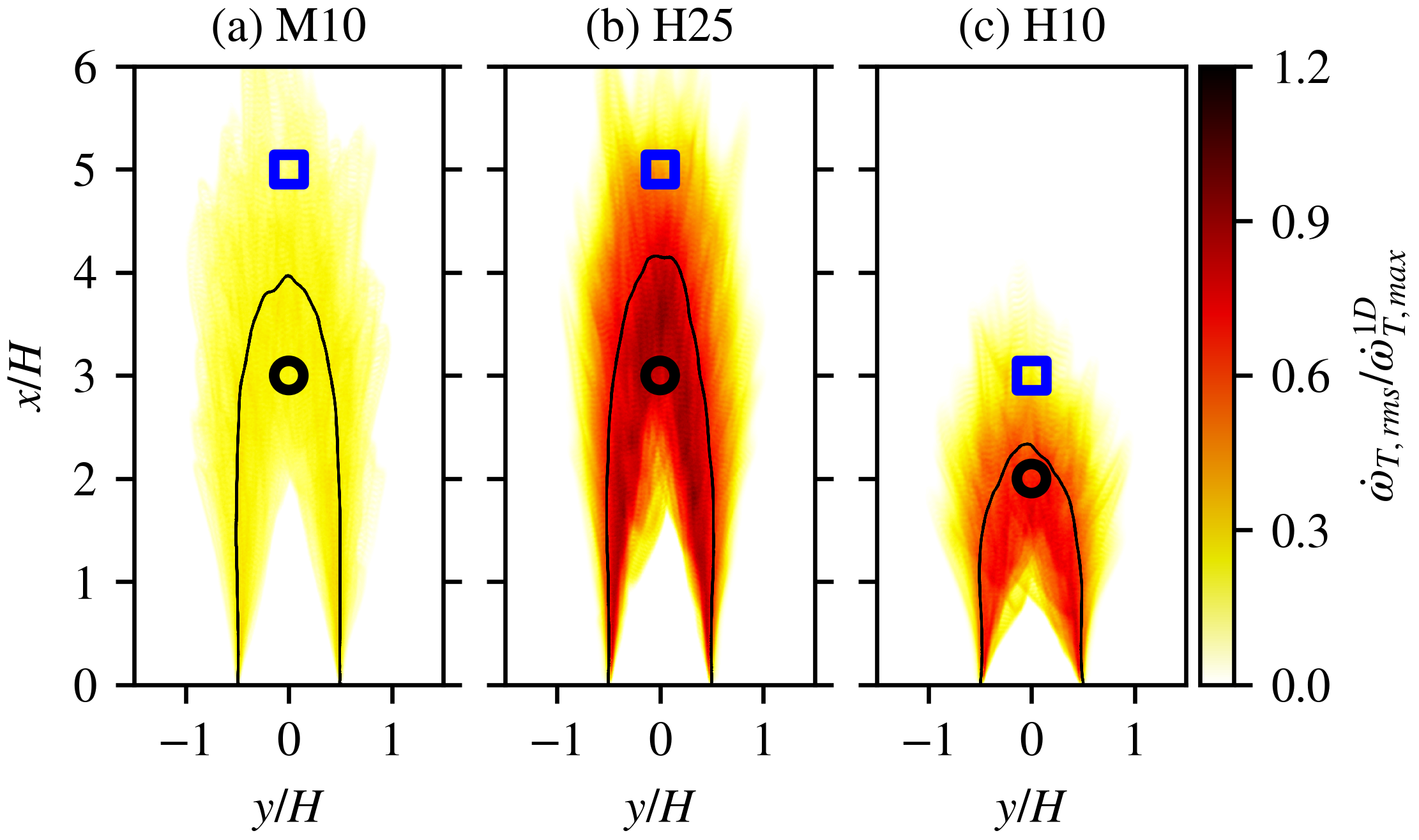}}
  \caption{Normalised temporal r.m.s. fields of the heat release rate $\dot{\omega}_{T,rms}/\dot{\omega}_{T,max}^{1D}$ in the $xy$ plane for the three cases.
  The black lines denote contours of mean progress variable $\tilde{C}=C^*$. The markers correspond to the positions of the probes in Fig.~\ref{fig:local_hr}.}
  \label{fig:hr_rms}
\end{figure}

\begin{figure}
  \centerline{\includegraphics[width=130 mm]{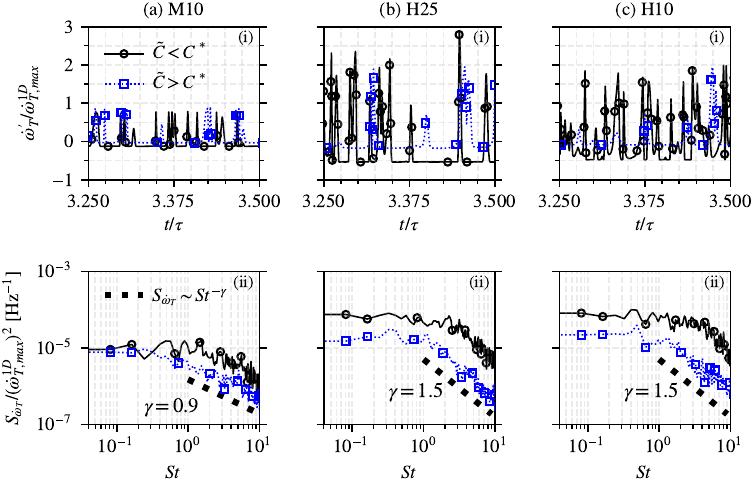}}
  \caption{Normalised temporal traces of local heat release rate fluctuations $\dot{\omega}_T'/\dot{\omega}_{T,max}^{1D}$ (i) and corresponding power spectral densities (ii) in the unburnt ($\tilde{C}<C^*$) and burnt ($\tilde{C}>C^*$) gas regions for the three cases.}
  \label{fig:local_hr}
\end{figure}

The differences between \ce{CH4} and \ce{H2} flames in the burning rate and HRR, which have been globally characterised in~\Cref{sec:flame_turbulence_int}, are here analysed locally to further understand the impact of TD effects on the combustion noise source term.
To this scope,~\Cref{fig:hr_rms} compares the normalised temporal r.m.s. of the local HRR $\dot{\omega}_{T,rms}/\dot{\omega}_{T,max}^{1D}$ in the $xy$ plane for the three flames, together with isocontours at $\widetilde{C}=C^*$ (in black) to highlight the mean flame position.
As expected, the \ce{CH4} flame (see~\Cref{fig:local_hr}(a)) shows an almost uniform value of $\dot{\omega}_{T,rms}$ across the entire flame brush, whose normalised intensity is close to the one of $\left|\nabla C\right|_{rms}/\left|\nabla C\right|_{max}^{1D}$ shown in~\Cref{fig:slot_strech_components_axis}(a).
This further sustains that, for this thermodiffusively stable flame, the variability in the local reactivity is caused by the sole action of turbulence on the flame surface, which, as discussed in~\Cref{sec:flame_turbulence_int,sec:flame_surface}, is dominated by annihilation events, especially at the flame tip.
On the other hand, the \ce{H2} flames show much stronger fluctuations of the local HRR in the inner region of the flame, i.e., for $\widetilde{C}<C^*$, which well relate to the enhancement of the flame surface generation mechanism associated with TD instabilities~\citep{berger2022intrinsic}.

Further insights can be obtained from~\Cref{fig:local_hr}, which reports, for each flame, the temporal traces of the local HRR fluctuations $\dot{\omega}_T'=\dot{\omega}_T-\overline{\dot{\omega}_T}$ (top), normalised by the maximum HRR in the 1-D laminar flame $\dot{\omega}_{T,max}^{1D}$ (see~\Cref{tab:lam_properties}), and the corresponding PSD (bottom) for two probes, identified by the markers in~\Cref{fig:hr_rms}, located in the unburnt gases region upstream of the mean flame position ($\widetilde{C}<C^*$) and in the downstream burnt gases region ($\widetilde{C}>C^*$).
It should be noted that $\dot{\omega}_T'$, being a small scale phenomenon, is highly intermittent.
Therefore, its value can be significantly higher than the time average $\overline{\dot{\omega}_T}$~\citep{swaminathan2011heat,liu2015modelling}.
Furthermore, $\dot{\omega}_T=0$ whenever the local equivalence ratio falls below the lean flammability limit, and $\dot{\omega}_T'$ is consequently limited, on the negative side, by the condition $\dot{\omega}_T'\geq-\overline{\dot{\omega}_T}$~\citep{kumar2025spectral}.

When comparing the temporal traces for the \ce{CH4} and \ce{H2} flames, significant differences arise.
Indeed, for the M10 case (see~\Cref{fig:local_hr}(a-i)), the two probes show comparable signals, in terms of shape of the distributions and amplitude of the oscillations, with the maximum positive value close to $\dot{\omega}_{T,max}^{1D}$.
The sole relevant difference is the larger number of bursts, denoting the presence of the flame at the given location within the considered time window, for the probe at $\widetilde{C}<C^*$ (circle-marked black solid line), which is to be expected given the higher availability of reactants in this zone.
Therefore, the \ce{CH4} flame locally displays a sort of \textit{on--off} behaviour: the HRR fluctuates with repeatable amplitude, and the main source of variability is represented by the occurrence of the bursts in time.
On the other hand, both the H25 (see~\Cref{fig:local_hr}(b-i)) and H10 (see~\Cref{fig:local_hr}(c-i)) flames consistently show an enhancement of HRR fluctuations, due to the action of TD effects (see~\Cref{sec:flame_turbulence_int}).
This is more accentuated in the probe located in the unburnt gas region ($\widetilde{C}<C^*$, square-marked blue dotted line), coherently with the results in~\Cref{fig:hr_rms}, while at $\widetilde{C}>C^*$ (circle-marked black solid line) this enhancement is less accentuated.
Furthermore, while the amplitude of the HRR bursts in the \ce{CH4} flame is almost constant, for the \ce{H2} a more accentuated variability is observed.
This indicates that, for the \ce{H2} flames, the local HRR is modulated not only by the sole effect of turbulence on the flame surface, which leads to the intermittent behaviour observed for the \ce{CH4} flame in~\Cref{fig:local_hr}(a-i), but also by the stretch-induced enhancement associated with TD effects.

These differences reflect in the corresponding spectra (see~\Cref{fig:local_hr}(ii)).
Coherently with the literature~\citep{kumar2025spectral}, all HRR spectra can be divided in two regions: a low-frequency ($St<St_c$) regime and a high frequency ($St>St_c$) one, where $St_c$ is a cut-off Strouhal number denoting the change in the HRR spectral behaviour: for $St<St_c$, the PSD is almost constant, while, for $St>St_c$, the PSD starts to decay following a power law (linear in logarithmic scale).
For the \ce{CH4} flame (see~\Cref{fig:local_hr}(a-ii)), the spectra at the two locations nearly collapse, especially in the low frequency region, while slight differences are observable in the high frequency one, due to a shift of $St_c$ toward lower values when moving downstream, from $St_c\approx1$ at $\widetilde{C}<C^*$ to $St_c\approx0.5$ at $\widetilde{C}>C^*$.
This is consistent with the reduced occurrence of burst events observed in the temporal traces (see~\Cref{fig:local_hr}(a-i)).
On the other hand, for the \ce{H2} flames (see~\Cref{fig:local_hr}(b-ii) and (c-ii)), a clear reduction in the value of the PSD is observable when moving downstream, consistently with the weaker amplitudes of the HRR bursts observed in the temporal signals at $\widetilde{C}>C^*$ (see~\Cref{fig:local_hr}(b-i) and (c-i)).
The value of $St_c$ increases with respect to the \ce{CH4} flame, being it equal to $St_c\approx 2$ for the probe at $\widetilde{C}<C^*$ and to $St_c\approx 1$ for the probe at $\widetilde{C}>C^*$, which means that the energetic low-frequency region is broader.
\citet{lapenna2024synergistic}, in their curvature-based analysis of turbulent flames, reported an increase in the cut-off wavenumber $k_c$ and a broader range of unstable wavenumbers when passing from thermodiffusively stable to unstable flames, thereby indicating that turbulence can interact synergistically with TD instabilities over a wider range of scales.
By relating spatial and temporal scales~\citep{pope_turbulent_2000}, the larger cut-off wavenumber $k_c$ observed by~\citet{lapenna2024synergistic} is coherent with the larger value of $St_c$ observed in the present work.

For $St>St_c$, on the other hand, the \ce{H2} flames show a steeper decay in the local HRR spectra, with the exponent of the roll-off power law passing from -0.9 for \ce{CH4} to -1.5 for \ce{H2}, thus indicating a suppression of strong HRR fluctuations at high frequencies.
This well relates to the observations made for laminar \ce{H2} flames by~\citet{berger2022intrinsic}, who showed, by means of linear stability analysis, that TD instabilities are favoured at large scales, while stabilization occurs at small ones.
This stabilization mechanism is consistent with the reduction in the flame speed upon annihilation observed by~\citet{talei2012parametric,talei2013direct}, and to the consequent reduced impact of flame annihilation phenomena on the high-frequency content discussed in~\Cref{sec:flame_surface}.
Finally, this spectral behaviour directly relates to the different roll-off in the acoustic spectra observed for \ce{CH4} and \ce{H2} flames in~\Cref{fig:pressure_spectra_farfield}.
Therefore, the multiscale action of TD instabilities on the local HRR distribution, leading to a transfer of HRR spectral energy from the small scales (at high frequency) to the large scales (at low frequency), combined with their influence on the flame surface area dynamics discussed in~\Cref{sec:flame_surface}, shape the combustion noise spectrum of lean premixed \ce{H2} flames, enhancing the low-frequency component and weakening the high-frequency one.
This represents the underlying mechanism explaining the differences in the acoustic spectra observed in~\Cref{sec:acoustic} between the \ce{CH4} and \ce{H2} flames.

\section{Modal analysis of noise generation mechanisms}
\label{sec:spod_noise}

To corroborate the analysis of the local temporal signals presented in the previous sections, SPOD of the pressure field is here adopted to identify the coherent structures associated with sound radiation across the frequency spectrum.
This allows for exploiting the spatial information contained in a broader dataset, thereby providing further insight into the spectral behaviour of the acoustic radiation and enabling a more detailed characterisation of the distinct noise generation mechanisms discussed earlier.
The modal decomposition is here performed using the Python library antares~\citep{antares}, with the SPOD implementation based on the work by~\citet{schmidt2020guide}.
The main features are recalled hereafter.

Each instantaneous spatial field $\boldsymbol{\phi}(\boldsymbol{x},t)$ is broken down in the sum of a temporal mean $\overline{\boldsymbol{\phi}}(\boldsymbol{x})$ and a fluctuating part $\boldsymbol{\phi}'(\boldsymbol{x},t)$.
The normalised pressure field $\boldsymbol{\phi}(\boldsymbol{x},t)=[p/p_{ref}](\boldsymbol{x},t)$ is extracted each $\Delta_t^{SPOD}/\tau=1.43\times10^{-3}$ and interpolated in a 3-D cylindrical domain, corresponding to the region of interest in~\Cref{fig:domain}, with grid size equal to 2.4 million tetrahedral cells.
A Hanning window with 75\% overlap is employed to prevent spectral leakage.
For each case, 2700 3-D solutions are considered, with $N_b = 11$ blocks of $N_{freq}=768$ instantaneous snapshots. 
This leads to a frequency resolution equal to $\Delta St\approx 0.05$ for all cases.
The use of different values of overlap, $N_b$ and $N_{freq}$ was not found to affect significantly the eigenvalues and the spatial patterns of the considered SPOD modes.

\begin{figure}
  \centerline{\includegraphics[width=130 mm]{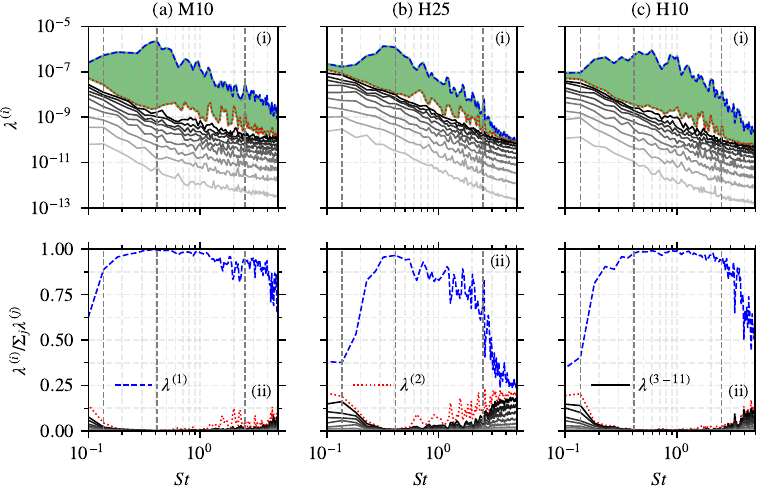}}
  \caption{Pressure SPOD eigenvalues $\lambda^{(i)}$ (i) and corresponding energy share $\lambda^{(i)}/\Sigma_j\lambda^{(j)}$ (ii) for the three cases. The green shaded area highlights the differences between the first two modes. The vertical dashed grey lines correspond to the modes shown in~\Cref{fig:spod_p_modes_low,fig:spod_p_modes_mid,fig:spod_p_modes_high}.}
  \label{fig:spod_pressure_spectrum}
\end{figure}

\Cref{fig:spod_pressure_spectrum} shows the eigenvalue $\lambda^{(i)}$ spectra for the three cases (top), together with the energy share $\lambda^{(i)}/\Sigma_j\lambda^{(j)}$ of each mode (bottom).
A low-rank behaviour is clearly observable, with the first mode dominating the spectral content, especially for $0.2\leq St\leq 1$.
A higher-rank behaviour is observed outside this range, especially for the \ce{H2} flames.

\begin{figure}
  \centerline{\includegraphics[width=130 mm]{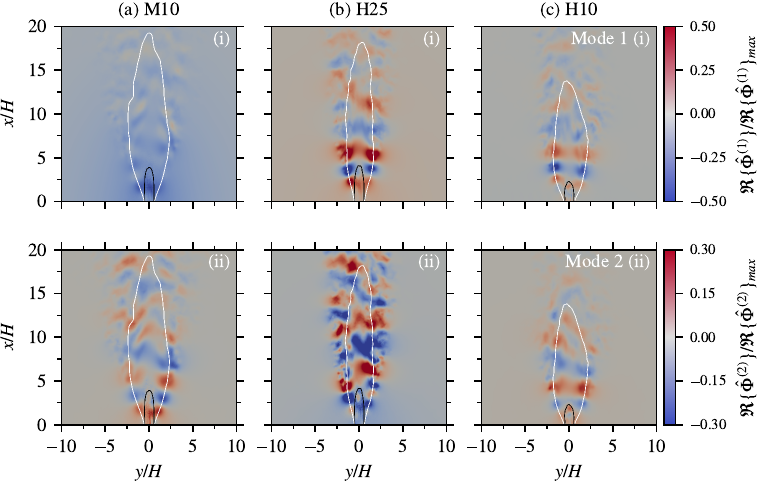}}
  \caption{Spatial distributions of the first (i) and second (ii) pressure SPOD mode at $St=0.15$ for the three cases. The black and white lines denote, respectively, contours of mean progress variable $\tilde{C}=C^*$ and of mean mixture fraction $\tilde{\xi}=0.5Y_{F,u}$.}
  \label{fig:spod_p_modes_low}
\end{figure}

Focusing on the low frequency acoustic radiation first,~\Cref{fig:spod_p_modes_low} reports the spatial distributions of the first two modes for the three cases at $St=0.15$ in the $xy$ plane, together with the mean flame and outer shear layer positions, identified, respectively, by the contours of mean progress variable $\widetilde{C}=C^*$ (black line) and of mean mixture fraction $\widetilde{\xi}=0.5Y_{F,u}$ (white line).
A clear difference arises between the \ce{CH4} and \ce{H2} flames in the first mode.
Indeed, for the former (see~\Cref{fig:spod_p_modes_low}(a-i)), no coherent structures are observed, but only a radiating mode originating from the flame.
A weak contribution is found in correspondence of the outer shear layer.
For the \ce{H2} flames (see~\Cref{fig:spod_p_modes_low}(b-i) and (c-i)), on the other hand, strong wavepacket structures are observable, coherent with the K-H instability deforming the outer shear layer~\citep{pickering2020lift,brouzet2020role,casel2022resolvent}.
These are stronger for the high-velocity H25 case (see~\Cref{fig:spod_p_modes_low}(a-i)), consistently with the hydrodynamic nature of this instability.
In the second mode (see~\Cref{fig:spod_p_modes_low}(ii)), \ce{H2} flames show analogous structures as in the first one, but in a weaker form, given the lower energy.
Moreover, the K-H wavepackets are present also in the \ce{CH4} flame, confirming the presence, albeit limited and only in the sub-leading mode, of these structures for the thermodiffusively stable case.
This sustains the analysis in~\Cref{sec:acoustic}, where it has been observed that instabilities at the outer shear layer are enhanced in the \ce{H2} flames.

\begin{figure}
  \centerline{\includegraphics[width=130 mm]{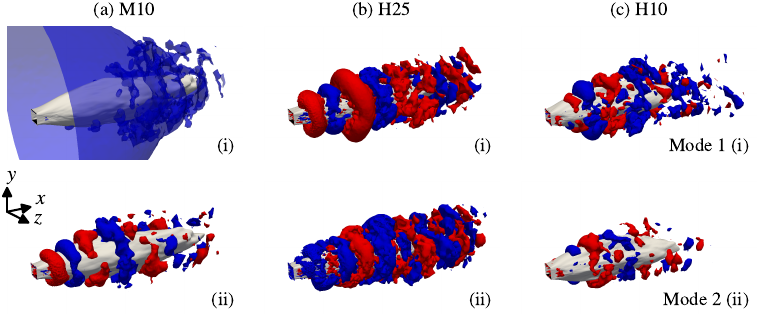}}
  \caption{Isosurfaces of the first (i) and second (ii) pressure SPOD mode at $St=0.15$ for the three cases taken at $-10\%$ (blue) and $+10\%$ (red) of their respective maximum value. The white isosurface corresponds to the mean mixture fraction $\tilde{\xi}=0.5Y_{F,u}$.}
  \label{fig:spod_p_modes_low_3d}
\end{figure}

The differences between the two fuels are further highlighted in~\Cref{fig:spod_p_modes_low_3d}, which shows contours of the two modes taken at $\pm10\%$ of the maximum value for the three cases.
The radiating nature of the first SPOD mode for the M10 case (see~\Cref{fig:spod_p_modes_low_3d}(a-i)) is evident, while the \ce{H2} flames show strong, spatially extended low-frequency coherent structures contouring the mean outer shear layer position (denoted by the white isosurface taken at $\widetilde{\xi}=0.5Y_{F,u}$), which are enhanced for the high-velocity H25 case.
Similar structures are present in the second mode for all flames (see~\Cref{fig:spod_p_modes_low_3d}(ii)), with a similar influence of the bulk velocity for \ce{H2} flames as for the first mode.
This indicates that, for the M10 case, flame-generated acoustic radiation, associated with the first mode, is the dominant noise generation mechanism at low $St$, while the outer shear layer provides a second-order contribution.
This is coherent with the strong difference in the energy share of the two modes observed in~\Cref{fig:spod_pressure_spectrum}(a-ii).
For the \ce{H2} flames, on the other hand, the K-H instability at the outer shear layer has a dominant contribution.
Energy is therefore transferred from the first to the second mode, consistently with the proximity in the energy share of the two modes observed at this frequency in~\Cref{fig:spod_pressure_spectrum}(b-ii) and (c-ii).
This difference can be related, on one hand, to the differences in the value of $T_{ad}$ and $\rho_b$ between the \ce{CH4} and \ce{H2} flames, since shear layer instabilities are less pronounced as the temperature ratio increases~\citep{trouve1988linear,schlimpert2016hydrodynamic,brouzet2020role}, and, on the other hand, to the stronger variability in the density field induced by \ce{H2} non-equidiffusion effects, previously observed in~\Cref{sec:acoustic}.

\begin{figure}
  \centerline{\includegraphics[width=130 mm]{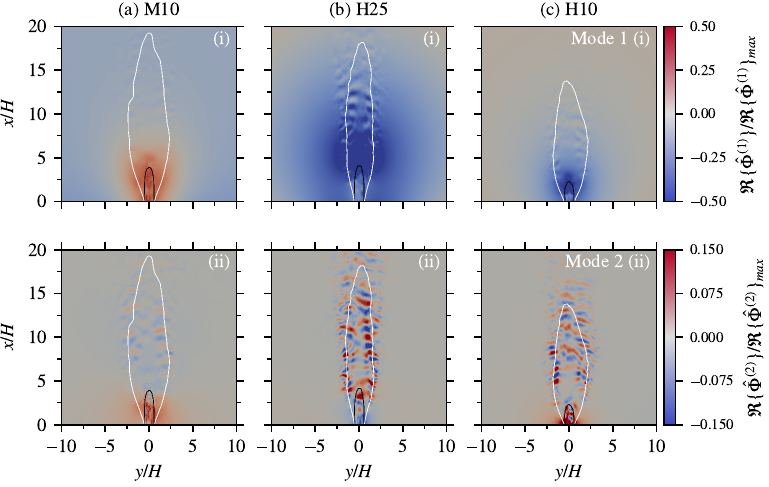}}
  \caption{Spatial distributions of the first (i) and second (ii) pressure SPOD mode at $St=0.4$ for the three cases. The black and white lines denote, respectively, contours of mean progress variable $\tilde{C}=C^*$ and of mean mixture fraction $\tilde{\xi}=0.5Y_{F,u}$.}
  \label{fig:spod_p_modes_mid}
\end{figure}

\Cref{fig:spod_p_modes_mid} reports the spatial modes in the $xy$ plane at $St=0.40$, i.e., in the mid-frequency range close to the peak of the acoustic spectrum (see~\Cref{fig:pressure_spectra_farfield}).
Here, the first mode (see~\Cref{fig:spod_p_modes_mid}(i)) has a radiating nature, centred in proximity of the mean flame position, for all cases, consistently with the dominance of flame-generated noise in this frequency region (see~\Cref{sec:acoustic}).
Still, a non-negligible trace of the wavepacket-like structures originated at the outer shear layer are visible in the \ce{H2} flames.
Therefore, for these flames, the low-frequency acoustic disturbance originated at the outer shear layer persists over a broader frequency range than in the \ce{CH4} case.
This is even more evident when looking at the second mode, for which no relevant coherent structures are present in the \ce{CH4} case, while clear, albeit weak, wavepackets are visible in correspondence of the outer shear layer for the \ce{H2} flames.
This difference can be associated to the finer-scale structures and enhanced mixing observed in this spatial region (see~\Cref{fig:slot_temp_rhoe_xy}), which translate in a persistence of these structures at higher frequencies. 

\begin{figure}
  \centerline{\includegraphics[width=130 mm]{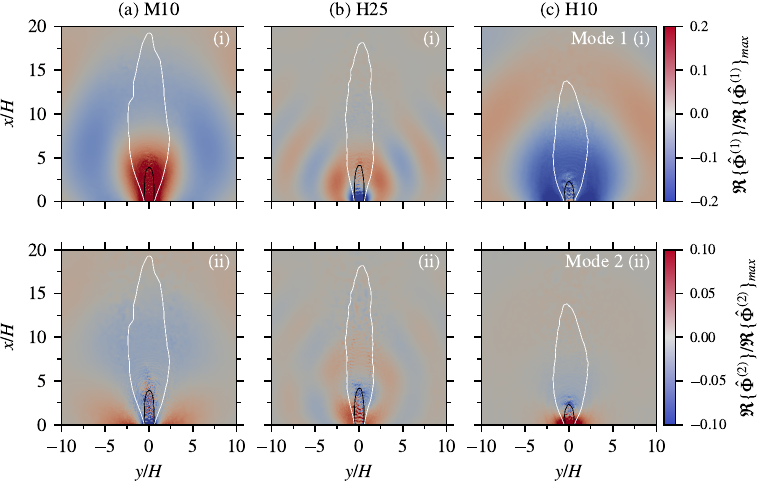}}
  \caption{Spatial distributions of the first (i) and second (ii) pressure SPOD mode at $St=2.5$ for the three cases. The black and white lines denote, respectively, contours of mean progress variable $\tilde{C}=C^*$ and of mean mixture fraction $\tilde{\xi}=0.5Y_{F,u}$.}
  \label{fig:spod_p_modes_high}
\end{figure}

Finally, the high frequency noise generation is considered by taking into account the spatial modes in the $xy$ plane at $St=2.5$, provided in~\Cref{fig:spod_p_modes_high}.
In this frequency regime, the pressure field is governed by the noise generated by the turbulent flame, while no outer shear layer structures are present either in the radiating first mode (see~\Cref{fig:spod_p_modes_high}(i)) or in the second one (see~\Cref{fig:spod_p_modes_high}(ii)).
The weak energy of the second mode indicates that the reduced energy share of the first mode at this value of $St$ observed in~\Cref{fig:spod_pressure_spectrum}(ii) is not associated to stronger higher-rank modes.
Rather, it is related to the insurgence of a weakly coherent regime and to the global weakening of the energy due to the high-frequency decay of the acoustic radiation.
Based on this modal decomposition, it is confirmed that the differences in the spectral decay observed in this frequency region are to be imputed to the differences in the HRR dynamics at the flame front discussed in~\Cref{sec:hrr_noise}, since the noise generated in the flame brush remains the dominant acoustic source at these frequencies.

\section{Conclusion}
\label{sec:conclusion}

The impact of thermodiffusive effects on the acoustic radiation of turbulent premixed slot jet flames in an open environment of ambient air at atmospheric conditions was investigated by means of Direct Numerical Simulations.
A thermodiffusively stable stoichiometric methane--air flame was compared with two lean hydrogen--air flames, selected to exhibit comparable laminar flame properties and subject to the same turbulence intensity.
The two hydrogen configurations differed in bulk jet velocity so as to reproduce either a turbulent flame brush length comparable to that of the methane flame or the same bulk velocity.

For both hydrogen flames, thermodiffusive effects were found to substantially alter the flame--turbulence interaction, directly affecting both turbulent flame surface dynamics and heat release rate fluctuations, and thereby modifying the far field acoustic radiation.
Increasing the bulk velocity enhanced the variability of the heat release rate and the production of flame surface, promoting stronger noise generation.
Conversely, when the bulk velocity was kept equal to that of the methane flame, the flame brush became shorter, leading to a shift of the combustion noise spectral peak towards higher frequencies.

Consistently with previous studies, the time derivative of the volume-integrated heat release rate was identified as the dominant far field noise source for all the configurations considered.
In addition, a theoretical framework extending the classical combustion noise flamelet theory to thermodiffusively unstable flames was proposed and validated, relating flame-generated noise to the time derivative of the turbulent flame surface area.
This result highlighted the role of flame stretch, showing that, all else being equal, the pressure fluctuations associated with combustion noise in lean premixed hydrogen flames are modulated by the stretch factor $I_0$, thereby supporting the noise-promoting role of thermodiffusive instabilities.

A systematic enhancement of the flame surface generation term was also observed in the hydrogen flames.
This mitigated the relative importance of destructive flame annihilation events, which are known to contribute to high-frequency acoustic radiation.
Furthermore, the analysis of local heat release rate spectra revealed stronger low-frequency fluctuations in the thermodiffusively unstable flames, together with a reduced high-frequency spectral content.
This was associated to the stabilising action of thermodiffusive instabilities at small scales.
From an acoustic standpoint, these features translated into two distinctive signatures of direct combustion noise in thermodiffusively unstable flames: an enhanced low frequency radiation and a steeper decay at high frequencies.

Finally, the influence of fuel properties on the pressure field in the shear layer between combustion products and ambient air was examined.
In the hydrogen flames, non-equidiffusion was found to induce stronger density gradients at this interface than in the methane flame, promoting  the Kelvin--Helmholtz instability mechanism.
SPOD of the pressure field revealed coherent low frequency structures in both hydrogen flames, more pronounced in the high velocity case, consistently reflecting the hydrodynamic nature of the instability.
Owing to the insurgence of finer scale structures in the thermodiffusively unstable cases, the wavepackets associated with the shear layer instability were also found to persist up to higher frequencies than in the methane flame.
Overall, these results suggest that thermodiffusive effects influence not only direct combustion noise, through their impact on flame surface dynamics and heat release rate fluctuations, but also indirect noise sources associated with excess density gradients.

\begin{bmhead}[Acknowledgements.]
The simulations were performed on the Luxembourg national supercomputer MeluXina. The authors gratefully acknowledge the LuxProvide teams for their expert support.
Additional computing resources and the related technical support used for this work have been provided by CRESCO/ENEAGRID High Performance Computing infrastructure and its staff~\citep{iannone2019cresco}. CRESCO/ENEAGRID High Performance Computing infrastructure is funded by ENEA, the Italian National Agency for New Technologies, Energy and Sustainable Economic Development and by Italian and European research programmes, see \url{http://www.cresco.enea.it/english} for information.
\end{bmhead}
\begin{bmhead}[Funding.]
The authors acknowledge EuroHPC JU for awarding the Project ID EHPC-REG-2023R03-184 (PROMETH2EUS) access to MeluXina at LuxProvide.
This work has been supported under the National Recovery and Resilience Plan (NRRP), Mission 4 Component 2 Investment 1.3 - Call for tender No. 1561 of October 11, 2022 of the Italian Ministry of University and Research, funded by the European Union - NextGenerationEU [Project code PE0000021, Concession Decree No. 1561 of October 11, 2022 adopted by the Italian Ministry of University and Research, Grant No. CUP - D93C22000900001, Project title ‘‘Network 4 Energy Sustainable Transition – NEST’’].
F.G.S. has received financial support for this research project by the French Ministry of Foreign Affairs and Europe and Campus France via a French Government Scholarship.
\end{bmhead}
\begin{bmhead}[Declaration of interests.]
The authors report no conflict of interest.
\end{bmhead}
\begin{bmhead}[Data availability statement.]
The data will be made available upon reasonable request.
\end{bmhead}
\begin{bmhead}[Author ORCIDs.]
F.G. Schiavone, https://orcid.org/0000-0003-2480-2596; G. Daviller, https://orcid.org/0000-0001-6644-9239; D. Laera, https://orcid.org/0000-0001-6370-4222
\end{bmhead}
\begin{bmhead}[Author contributions.]
F.G.S. performed the numerical simulations and the postprocessing, analysed and cured data, and wrote the original draft.
G.D. and D.L. were responsible for supervision.
D.L. acquired funding.
All authors conceptualised the work and contributed to reviewing and editing the manuscript.
\end{bmhead}

\appendix
\begin{appen}

\section{Definition of the reference acoustic pressure}\label{app:pressure}
The acoustic power $\mathcal{P}$ radiated from an open turbulent premixed flame scales with the product of the squares of the flame thermal power $P_{th}$ and of a characteristic frequency $f_{char}=U_B/L_f$~\citep{rajaram2006premixed,candel2009flame}.
The following relation may be established:

\begin{equation}
    \mathcal{P}\approx \frac{(\gamma-1)^2}{4\pi\rho_\infty c_\infty^5}f_{char}^2P_{th}^2=\frac{(\gamma-1)^2}{4\pi\rho_\infty c_\infty^5}\left(\frac{U_B}{L_f}\right)^2\left(Y_{F,u}\rho_uA_{ch}U_Bh_{lv}\right)^2=\frac{(\gamma-1)^2}{4\pi\rho_\infty c_\infty^5}\left(\frac{Y_{F,u}\rho_u1.5H^2U_B^2h_{lv}}{L_f}\right)^2.
\end{equation}
Recalling~\Cref{eq:noise_hrr} and the definition of the acoustic power:

\begin{equation}
    \mathcal{P}=\frac{\overline{p'^2}}{\rho_\infty c_\infty}4\pi\left|\boldsymbol{x}\right|^2,
\end{equation}
an expression for $\overline{p'}$ can then be derived:

\begin{equation}
    \overline{p'}=\sqrt{\mathcal{P}\frac{\rho_\infty c_\infty}{4\pi\left|\boldsymbol{x}\right|^2}}\approx\sqrt{\frac{(\gamma-1)^2}{4\pi\cancel{\rho_\infty} c_\infty^{\cancel{5}4}}\left(\frac{Y_{F,u}\rho_u1.5H^2U_B^2h_{lv}}{L_f}\right)^2\frac{\cancel{\rho_\infty} \cancel{c_\infty}}{4\pi\left|\boldsymbol{x}\right|^2}}.
\end{equation}
By neglecting the constant term $1.5^2$ and assuming a reference distance $\left|\boldsymbol{x}\right|=H$, then:

\begin{equation}
    \overline{p'}\approx \sqrt{\frac{(\gamma-1)^2}{4\pi c_\infty^4}\left(\frac{Y_{F,u}\rho_uH^2U_B^2h_{lv}}{L_f}\right)^2\frac{1}{4\pi H^2}}=\frac{(\gamma-1)Y_{F,u}\rho_uH^{\cancel{2}}U_B^2h_{lv}}{4\pi c_\infty^2L_f\cancel{H}},
\end{equation}
which corresponds to the expression of $p_{ref}$ given in~\Cref{eq:p_ref}.

\section{Spectral analysis of turbulence}\label{app:turbulence}

For the considered configurations, the turbulence injected in the unburnt mixture is the dominant source of velocity fluctuations affecting the flames~\citep{coulon2023direct,gaucherand2024dns}.
Nevertheless, additional perturbations in the velocity field originate from the turbulence developed in the wall-bounded channel flow upstream of the flame, leading to a deviation of the turbulent field from the injected one.
To quantify the evolution of turbulence in the inlet channel,~\Cref{fig:turb_spectra} reports, for the three cases, the power spectral densities of the turbulent kinetic energy $S_{TKE}$ at different positions, with $S_{TKE} = (S_u+S_v+S_w)/2$, where $S_u$, $S_v$ and $S_w$ are the power spectral densities of the velocity components in the $x$, $y$ and $z$ directions, respectively.
The turbulence decays quite slowly, and a similarity can be observed among the three cases in the shape and magnitude of the power spectral densities, especially for the probes in proximity of the flame.
Due to the high viscous dissipation, a lower power spectral density is observed for the probes near the outer shear layer, as well as for the farthest probe in the axial direction of the H10 case, being this latter fully in the combustion products (see~\Cref{fig:turb_dissipation}).

\begin{figure}
  \centerline{\includegraphics[width=130mm]{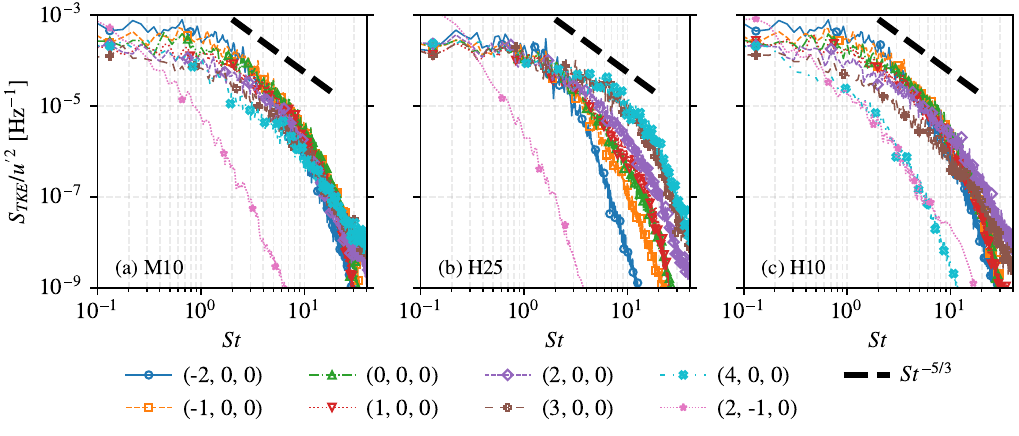}}
  \caption{Power spectral density of the turbulent kinetic energy $S_{TKE}$, normalised by $u'=2.5$~m~s$^{-1}$, retrieved from different probes for the three cases. The coordinates of the probes are given as $(x/H, y/H, z/H)$. The line at $St^{-5/3}$ is added for reference.}
  \label{fig:turb_spectra}
\end{figure}

Moreover, shear-induced turbulence is developed in the inner shear layers surrounding the fresh gases and outer shear layers surrounding the burnt gases.
The former induces an increase in viscosity due to the high temperature increase generated across the flame front, while the latter is a non-reacting interface, with no-relevant influence on the flame dynamics.
Both add significant dissipation to the turbulent structures, as it can be observed from~\Cref{fig:turb_dissipation}, which reports the Favre-averaged turbulence dissipation rate $\widetilde{\varepsilon}$, with the average performed in time over a windows of duration $\tau$ and in the span-wise direction $z$ for the whole extension of the slot, {\it i.e.} between $z/H=-0.75$ and $z/H=0.75$.
The contours of progress variable $\widetilde{C}=C^*$ (in white) and of $\widetilde{\xi}=0.5Y_{F,u}$ (in red) are also reported, to highlight the position of the inner and outer shear layers, respectively.
The markers correspond to the positions of the probes of the spectra of~\Cref{fig:turb_spectra}.

\begin{figure}
  \centerline{\includegraphics[height=55mm]{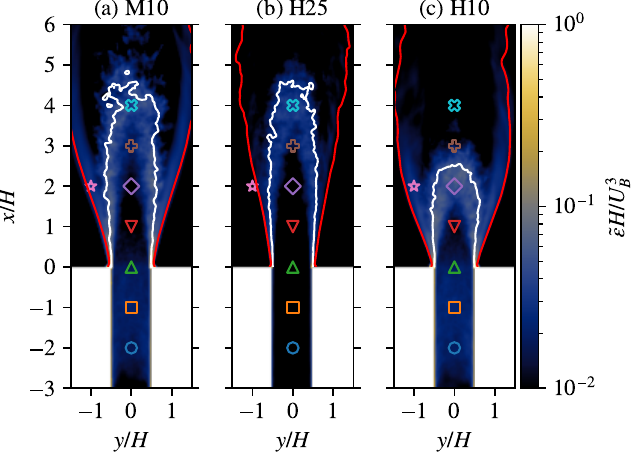}}
  \caption{Normalised turbulence dissipation rate $\tilde{\varepsilon}H/U_B^3$ and contours of progress variable $\tilde{C}=C^*$ (in white) and mixture fraction $\tilde{\xi}=0.5Y_{F,u}$ (in red). The markers denote the position of the probes of~\Cref{fig:turb_spectra}.}
  \label{fig:turb_dissipation}
\end{figure}

\section{Reference parameters for the transport model}\label{app:transport}

Since the mixtures are perfectly premixed and mainly composed of air, the molecular viscosities $\mu$ for the multi-species flows are assumed to be dependent only on the temperature and not on the gas composition, using the power law:

\begin{equation}
    \mu = \mu_{ref}\left(\frac{T}{T_{ref}}\right)^\beta,
    \label{eq:mol_visc}
\end{equation}
The reference parameters $\mu_{ref}$, $T_{ref}$, and $\beta$ to compute the molecular viscosity (see~\Cref{eq:mol_visc}) are reported in~\Cref{tab:viscosity}.
These are determined using least squares optimisation to fit the viscosity computed with the law of~\citet{wilke1950viscosity} in 1-D flame calculations at the chosen operating conditions.

\begin{table}
  \begin{center}
    \def~{\hphantom{0}}
    \begin{tabular}{lccc}
      Fuel & $\mu_{ref}$ [kg m$^{-1}$ s$^{-1}$] & $T_{ref}$ [K] & $\beta$\\[3pt]
      \ce{CH4} & $1.81\times10^{-5}$ & 300 & 0.68\\  
      \ce{H2}  & $8.06\times10^{-5}$ & 2645 & 0.65\\
    \end{tabular}
    \caption{Reference parameters used in the power law to evaluate the molecular viscosity.}
    \label{tab:viscosity}
  \end{center}
\end{table}

The diffusion coefficients $D_k$ of each species $k$ are calculated from the Schmidt numbers $Sc_k$:

\begin{equation}
    D_k = \frac{\mu}{\rho Sc_k},
\end{equation}
while the thermal conductivity $\lambda$ is evaluated from the Prandtl number $Pr$ and specific heat capacity $\overline{C}_p$ of the mixture:

\begin{equation}
    \lambda = \frac{\overline{C}_p\mu}{Pr}.
\end{equation}
The values of $Sc_k$ and $Pr$, reported in~\Cref{tab:transport}, are chosen to fit the detailed transport coefficients in the targeted conditions.

\begin{table}
  \begin{center}
    \def~{\hphantom{0}}
    \begin{tabular}{lccccccccccccc}
      \multirow{2}{*}{Fuel}   & \multirow{2}{*}{$Pr$} & \multicolumn{12}{c}{$Sc_k$}\\
       & & \ce{CH4} & \ce{CO2} & \ce{CO} & \ce{H2} & \ce{H2O} & \ce{H} & \ce{OH} & \ce{HO2} & \ce{H2O2} & \ce{O2} & \ce{O} & \ce{N2} \\[3pt]
      \ce{CH4} & 0.68 & 0.68 & 0.95 & 0.75 & - & 0.54 & - & - & - & - & 0.74 & - & 0.73\\
      \ce{H2}  & 0.66 & - & - & - & 0.23 & 0.58 & 0.14 & 0.53 & 0.80 & 0.81 & 0.80 & 0.52 & 0.91\\
    \end{tabular}
    \caption{Prandtl and Schmidt numbers adopted to compute diffusivities.}
    \label{tab:transport}
  \end{center}
\end{table}

\section{Influence of mesh size}\label{app:mesh}
A study of influence of the mesh size is performed for the H25 case, since it is prone to thermodiffusive instabilities and features the smallest Kolmogorov length scale (see~\Cref{tab:flow_params}).
To this scope, two additional simulations have been performed by considering a grid resolution in the reacting region equal to 80~$\mu$m and 60~$\mu$m, leading to mesh sizes equal, respectively, to 270 million and 490 million cells.
The flame structure statistics are represented in~\Cref{fig:mesh_influence} by the conditional mean and standard deviation of the hydrogen reaction rate $\dot{\omega}_{H_2}$, normalised by the maximum value in the corresponding 1-D laminar unstretched flame $\dot{\omega}_{H_2,max}^{1D}$, related to the progress variable $C$, defined as in~\Cref{eq:progvar}.
The statistics are not affected by successive refinements, demonstrating that $\Delta_x=100$~$\mu$m is sufficient, coherently with the previous work by~\citet{male2025hydrogen} for turbulent premixed hydrogen--air flames under analogous operating conditions.

\begin{figure}
  \centerline{\includegraphics[height=45mm]{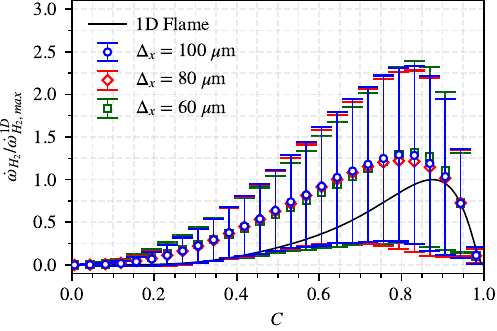}}
  \caption{Conditional mean (symbol) and standard deviation (error bar) of the normalised hydrogen reaction rate $\dot{\omega}_{H_2}/\dot{\omega}_{H_2,max}^{1D}$ with respect to the progress variable $C$ for different mesh resolutions. The mean hydrogen reaction rate for the corresponding 1-D laminar unstretched flame (black line) is added for reference.}
  \label{fig:mesh_influence}
\end{figure}

\section{Definition of flame stretch and speed}\label{app:stretch}
The total flame stretch $\kappa$, measured on the flame surface at $C=C^*$, is defined as:

\begin{equation}
    \kappa=\frac{1}{A}\frac{\text{d}A}{\text{d}t},
\end{equation}
where $A$ is the local flame surface area element.
The stretch is the sum of two contributions: the tangential strain rate $\kappa_s$ and the curvature term $\kappa_c$~\citep{candel1990flame}:

\begin{equation}
    \kappa = \kappa_s + \kappa_c = \underbrace{-\boldsymbol{n}\boldsymbol{n}:\nabla\boldsymbol{u}+\nabla\cdot\boldsymbol{u}}_{\kappa_s}+\underbrace{2\left.S_d\right|_{C=C^*}K}_{\kappa_c},
\end{equation}
where $K=(\nabla\cdot\boldsymbol{n})/2$ is the (geometrical) flame curvature, $\boldsymbol{n}=-\nabla C/\left|\nabla C\right|$ is the normal vector to the flame front, pointing towards the fresh gases, and $\boldsymbol{u}$ is the flow velocity.
Consequently, $K>0$ corresponds to a flame element curved convexly towards the fresh gases, while the opposite holds for $K<0$.
Finally, the curvature term $\kappa_c$ can be decomposed in its positive and negative components as~\citep{berger2022synergistic,coulon2023direct}:

\begin{equation}
    \kappa_c^+=\int_0^{+\infty}\kappa_c\mathcal{P}(\kappa_c)\text{d}\kappa_c\,\,\,\text{and}\,\,\,\
    \kappa_c^-=\int_{-\infty}^0\kappa_c\mathcal{P}(\kappa_c)\text{d}\kappa_c,
\end{equation}
where $\mathcal{P}(\kappa_c)$ is the probability to find the curvature value $\kappa_c$.
It is noted that $\kappa_c^-$ corresponds to regions where the flame propagates in the direction where the centre of curvature is located, while the opposite holds for $\kappa_c^+$.

The displacement speed $\left.S_d\right|_{C=C^*}$ is defined as the difference of the absolute flame speed $\left.S_a\right|_{C=C^*}=\boldsymbol{w}\cdot\boldsymbol{n}$, where $\boldsymbol{w}$ is the velocity vector of a point on the flame surface defined by the condition $C=C^*$, and of the normal component of the flow velocity vector $\boldsymbol{u}\cdot\boldsymbol{n}$~\citep{poinsot2005theoretical}:

\begin{equation}
    \left.S_d\right|_{C=C^*}=\left.S_a\right|_{C=C^*}-\boldsymbol{u}\cdot\boldsymbol{n}=\boldsymbol{w}\cdot\boldsymbol{n}-\boldsymbol{u}\cdot\boldsymbol{n}=\left.\frac{1}{\left|\nabla C\right|}\frac{\partial C}{\partial t}\right|_{C=C^*}+\left.\boldsymbol{u}\cdot\frac{\nabla C}{\left|\nabla C\right|}\right|_{C=C^*}.
\end{equation}
The value of $\left.S_d\right|_{C=C^*}$ is dependent on the considered $C=C^*$ isolevel considered.
In particular, for high values of $C^*$, the measure is altered due to gas acceleration through the flame front.
For a meaningful comparison between values defined on different isosurfaces, the density-weighted displacement speed is introduced~\citep{giannakopoulos2015consistent}:

\begin{equation}
    S_d=\frac{\left.\rho S_d\right|_{C=C^*}}{\rho_u}.
\end{equation}

Finally, the average of a quantity $\varsigma$ on the surface $S$ identified by the condition $C=C^*$ is here defined as:

\begin{equation}
    \langle \varsigma\rangle_S=\frac{\int_S\varsigma dA}{\int_SdA}=\left.\frac{\int_V\varsigma\left|\nabla C\right|dV}{\int_V\left|\nabla C\right|dV}\right|_{C=C^*}.   
\end{equation}
It is recalled that the surface-averaged total stretch is related to the variation of the turbulent flame surface area $A_T$~\citep{berger2024effects}.

\end{appen}

\bibliographystyle{jfm}
\bibliography{jfm_biblio}

\end{document}